%% file: smc.tex
\RequirePackage{rotating}
\documentclass[iop]{emulateapj} 
\usepackage{xspace}
\usepackage{natbib}
\usepackage{hhline}
\usepackage{amsmath}
\usepackage{dcolumn}
\usepackage{verbatim}
\usepackage{calc}
\usepackage{color}
\usepackage{booktabs}
\usepackage{longtable}
\usepackage[bookmarks]{hyperref}
\usepackage[numbered]{bookmark}
\usepackage{rotating}
\usepackage{paralist}
\usepackage[multidot]{grffile}
\usepackage{tabto}
\newcolumntype{.}{D{.}{.}{-1}}

\newcommand{\rosat}{{\it ROSAT}\xspace}
\newcommand{\chandra}{{\it Chandra}\xspace}

\newcommand{\swift}{{\it Swift}\xspace}
\newcommand{\xmm}{{\it XMM-Newton}\xspace}
\newcommand{\asca}{{\it ASCA}\xspace}
\newcommand{\rxte}{{\it RXTE}\xspace}
\newcommand{\integral}{{\it INTEGRAL}\xspace}

\newcommand{\sS}[1]{\mbox{$\rm{}^{#1}$}}
\newcommand{\Ss}[1]{\mbox{$\rm{}_{#1}$}}
\newcommand{\nep}[2]{\mbox{${#1}$$\times$${10}^{#2}$}}

\newcommand{\Ms}{\mbox{$M_{\odot}$}\xspace}

\newcommand{\nH}{\mbox{$N_{\mbox{\scriptsize H}}$}\xspace}

\newcommand{\Deg}{\mbox{$^\circ$}\xspace}

\newcommand{\lcgs}{\mbox{erg s\sS{-1}}\xspace}

\newcommand{\Zs}{\mbox{$Z_{\odot}$}\xspace}

\definecolor{ao}{rgb}{0.0, 0.5, 0.0}
\definecolor{amber}{rgb}{1.0, 0.49, 0.0}
\definecolor{ballblue}{rgb}{0.13, 0.67, 0.8}
\definecolor{bleudefrance}{rgb}{0.19, 0.55, 0.91}
\definecolor{brandeisblue}{rgb}{0.0, 0.44, 1.0}

\begin{document}

%%text_start
\title[SMC Pulsars]{
Deep \chandra Survey of the Small Magellanic Cloud. \\
II.~Timing Analysis of X-ray Pulsars
}

% a separate paper for variable sources

%\slugcomment{for ApJ}

\author{
JaeSub Hong\altaffilmark{1*},
Vallia Antoniou\sS{1},
Andreas Zezas\sS{2,3,1},
Frank Haberl\sS{4},
Manami Sasaki\sS{5}, 
Jeremy Drake\sS{1},
Paul Plucinsky\sS{1},
Silas Laycock\sS{6} \\
\textnormal{\footnotesize
\sS{1}{Harvard-Smithsonian Center for Astrophysics, 60 Garden St., Cambridge, MA 02138, USA: jaesub@head.cfa.harvard.edu} \\
\sS{2}{Foundation for Research and Technology-Hellas, 71110 Heraklion, Crete, Greece} \\
\sS{3}{Physics Department \& Institute of Theoretical \& Computational Physics, University of Crete, 71003 Heraklion, Crete, Greece}  \\
\sS{4}{Max-Planck-Institut f{\"u}r extraterrestrische Physik, Giessenbach stra\ss e, 85748 Garching, Germany} \\
\sS{5}{University of Erlangen-Nurnberg/Remeis Sternwarte}  \\
\sS{6}{Department of Physics, University of Massachusetts Lowell, MA 01854, USA} \\
}
}
%\altaffiltext{*}{Send requests to J. Hong at jaesub@head.cfa.harvard.edu}
%\altaffiltext{1}{Harvard-Smithsonian Center for Astrophysics, 
%60 Garden St., Cambridge, MA 02138 }

\begin{abstract}

We report the timing analysis results of X-ray pulsars from a recent
deep \chandra survey of the Small Magellanic Cloud (SMC).  We have
analyzed a total exposure of 1.4 Ms from 31 observations over a
1.2 deg\sS{2} region in the SMC under a \chandra X-ray Visionary Program.
Using the Lomb-Scargle and epoch folding techniques, we have detected
periodic modulations from 20 pulsars and a new candidate pulsar.
The survey also covers 11 other pulsars with no clear sign of periodic
modulation.
The 0.5--8 keV X-ray luminosity ($L_X$) of the pulsars ranges from 10\sS{34} to
10\sS{37} erg s\sS{-1} at 60 kpc.  All the \chandra sources with $L_X$
$\gtrsim$ \nep{4}{35}~\lcgs exhibit X-ray
pulsations.  The X-ray spectra of the SMC pulsars (and high mass
X-ray binaries) are in general harder than those of the SMC
field population.  All but SXP~8.02 can be fitted by
an absorbed power-law model with a photon index of $\Gamma$ $\lesssim$ 1.5.
The X-ray spectrum of the known magnetar SXP~8.02 is better fitted
with a two-temperature blackbody model.  
Newly measured pulsation periods of SXP~51.0, SXP~214 and SXP~701 are significantly
different from the previous \xmm and \rxte measurements. 
This survey provides a rich data set for energy-dependent pulse
profile modeling.  Six pulsars show an almost eclipse-like dip in the
pulse profile.  Phase-resolved spectral analysis reveals diverse
spectral variation during pulsation cycle: e.g., for an absorbed
power-law model, some exhibit an (anti)-correlation between absorption
and X-ray flux, while others show more intrinsic spectral variation
(i.e., changes in photon indices).

\end{abstract}

\section{Introduction} \label{s:intro}

The Small Magellanic Cloud (SMC) is a dwarf irregular satellite
of the Milky Way. Located nearby 
\citep[$\sim$ 60 kpc,][]{Hilditch05}
with a relatively unobstructed view (\nH$\sim$10\sS{21} cm\sS{-2}),  
the SMC is an ideal place to study stellar evolution
as it is experiencing an era of intense star formation 
\citep[0.05--0.4 \Ms yr\sS{-1},][]{Harris04,Shtykovskiy05}, which is likely
triggered by the tidal interaction with the Large Magellanic
Cloud (LMC) and our Galaxy \citep{Zaritsky04}.
The SMC harbors a large population of young X-ray binaries (XRBs),
dominantly in forms of High Mass XRBs (HMXBs) with Be star companions.
Thus, the SMC hosts an usually high number of Be-XRBs 
\citep[e.g.,][]{Coe15,Haberl16,Antoniou16}, which is likely
linked to the recent star formation episode
and low metallicities in the region \citep[e.g.,][]{Antoniou10}.

In Be-XRBs, the compact object, a spinning neutron star (NS),
accretes the material from the circumstellar disk
of the Be star, triggering bright X-ray outbursts with typical X-ray
luminosity of $\sim$ 10\sS{36-37} \lcgs.
Long-term monitoring surveys of the SMC with
\rxte \citep[e.g.,][]{Laycock05,Galache08} detected more than 50
pulsars during
outbursts, predominantly in the SMC Bar.  Observations
with \xmm \citep[e.g.,][]{Sturm13} and \chandra
\cite[e.g.,][]{Laycock10} have also detected dozens of new pulsars in
the region, and also extended the detection of XRBs to the quiescent regime
($\sim$10\sS{33} \lcgs).

In order to establish a full census of all active accreting binaries down
to $L_X$~$\sim$~10\sS{32}~\lcgs, a deep \chandra survey of the 11 fields in
the SMC was recently conducted under a \chandra X-ray Visionary Program
(PI.~A.~Zezas).  The goal of the survey is to acquire the deepest X-ray
luminosity function (XLF) in the region, and thus to measure the formation
efficiency of XRBs as a function of age, and the evolution of the XLF,
which can be used to constrain the XRB population synthesis models.
The details of the survey program along with the full source catalog
are found in Antoniou et al.~(2017, in preparation). 

These deep \chandra observations provide the spectral and timing
information for the SMC X-ray pulsars (SXPs) with unprecedented
high precision thanks to the low background enabled by the superb angular
resolution of the \chandra X-ray observatory.  In this paper, we report
the X-ray timing analysis results of the SXPs from our \chandra survey.
In \S\ref{s:obs} we outline the observations and analysis pipeline
for source search and aperture photometry. In \S\ref{s:timing} we describe
the procedures for timing analysis and
the phase-resolved spectral analysis using spectral model
fitting, energy quantiles and energy-versus-phase diagrams.
In \S\ref{s:pulsars} we present the pulsation search results
and review the properties of each pulsar that
exhibited X-ray pulsations during our survey.
In \S\ref{s:discussion} we compare the properties of pulsars with
those of other HMXBs and general field sources in the SMC, and
summarize the interesting features of selected pulsars.

%%text_stop
\begin{table*}
\tiny
\begin{minipage}{0.99\textwidth}
\caption{\chandra Observations of the SMC}
\input{tab1.tex}
\label{t:obs}
Notes. 
(1) Deep Field (DF) designation ID. 
(2) Stacked observations for timing analysis.
(3) Good time intervals (GTIs).
(4) The number of sources detected by the {\it wavdetect} routine in the 0.5--7 keV band \citep{Freeman02}.
(5) Sources with $\geq$ 100 net counts in the 0.3--8 keV band.
(6) The number of periodic sources. The number in the parenthesis indicates
duplicate sources detected in the other repeated observations.
\end{minipage}
\end{table*}
%%text_start

%%text_stop
\begin{table}
\tiny
\begin{minipage}{0.49\textwidth}
\caption{Analysis of Five Stacked Fields in the SMC}
\input{tab2.tex}
\label{t:obs_stacked}
See the notes in Table~\ref{t:obs}.
\end{minipage}
\end{table}
%%text_start
\section{Observations and Data Processing} \label{s:obs}

We have conducted a \chandra survey of seven and four fields in the Bar and
Wing regions of the SMC, respectively, from 2012 December to 2014 March
with a total exposure of 100 ks per each field.  Back in 2001 and 2006,
three other fields in the SMC Bar were observed with \chandra also for 100
ks each.  We have analyzed all 14 deep \chandra fields for the
timing analysis of the SMC pulsars.  Table~\ref{t:obs} summarizes the
deep \chandra observations of the SMC fields analyzed 
in this paper and the source count in each field.  

Each deep field overlaps with its neighboring fields to varying degrees.
Except for the observation of the NGC346 field (Obs.~ID 1881), the 100 ks exposure
of each field was broken into two (and in a few cases three) segments
in part due to the
observing constraints of \chandra. 
The sensitivity of pulsation detection improves when combining multiple
observations unless they are far apart in time, which can introduce
phase mixing.  Thus, when a field is repeatedly
observed within a week or two, we stack the observations for
pulsation search.  For instance, Deep Field 3 (DF3) was observed on
three separate occasions (Obs.~IDs 14666, 15499, and 16490) and for
the timing analysis, we stack the last two of these observations (Obs.~IDs
15499 and 16490) after separate searches in individual observations.
Table~\ref{t:obs_stacked} lists the run-down source counts in the
stacked fields.

For the analysis of both the individual observations and the stacked
data we employed the latest version of the X-ray analysis pipeline developed
for the \chandra Multi-wavelength
Plane survey \citep[ChaMPlane;][]{Grindlay05, Hong12}.  The ChaMPlane
pipeline is independent of the analysis pipeline used in Antoniou
et al.~(2017, in preparation), but the former 
was chosen for timing analysis because
of its proven maturity of periodicity search and subsequent
analysis tools \citep[e.g.,][]{Hong09, Laycock10, Hong12b}. 
Since a reliable timing analysis requires at least moderately
bright sources with $\gtrsim$ 100 net counts, the minor differences
between the two pipelines that only matter for relatively faint
sources ($\lesssim$ 20 net counts) do not change the outcome of this analysis.

The latest version of the ChaMPlane pipeline is based on CXC CIAO version
4.6 or later.  In summary, we searched for point sources using {\it
wavdetect} \citep{Freeman02} in the 0.5--7 keV band\footnote{For source
detection we chose a section of the broad band (0.3--8 keV) used
for the photometry, where the \chandra ACIS-I chips are most sensitive. 
This is to detect all the bright sources efficiently regardless of their spectral
hardness without need for additional source searches in separate soft
and hard bands.  } and performed the aperture photometry in multiple
energy bands to extract basic X-ray properties such as net counts, energy
quantiles, X-ray fluxes and luminosities.  For bright sources with $\ge$
100 net counts in the 0.3--8
keV band, we searched for periodic X-ray modulation (\S\ref{s:timing}). 
We have detected 2339 X-ray point sources from 31 individual
pointings, and have found periodic modulations from 18 sources
among 244 bright sources.  In the stacked data sets, we have detected
458 X-ray point sources and found 10 periodic sources among 52 bright
sources.  The last three columns in Tables~\ref{t:obs} and
\ref{t:obs_stacked} summarize source counts from individual and stacked
data sets, respectively. Note that source counts in Table~\ref{t:obs}
include many duplicate sources from the repeated observations.

\begin{figure*} 
\begin{center}
\includegraphics*[width=0.90\textwidth,clip=true]{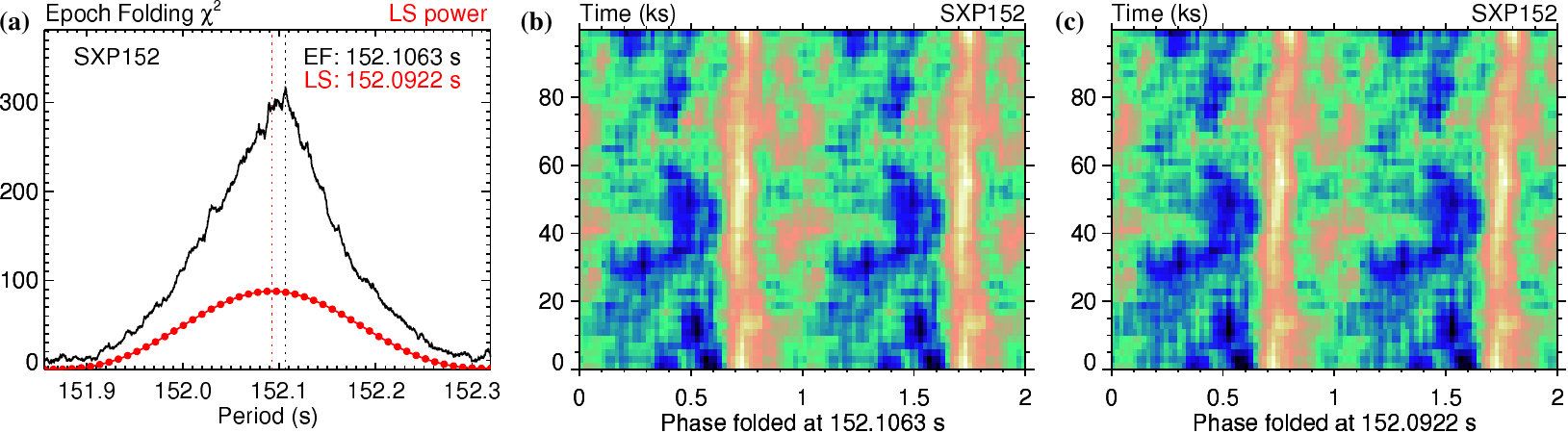}
\caption{Searching for the true modulation period of SXP~152 
between the peak periods of the refined LS (red) and EF (black)
periodograms (a) using the time versus phase diagrams (b and c).
Darker (bluer) and lighter (yellower) colors in the time versus phase
diagrams indicate relatively high and low counts, respectively. 
Unlike the peak period of the EF periodogram (b), the peak period of
the LS periodogram (c) shows a noticeable phase drift: 
the minimum is near phase $\sim$ 0.7 at T $\sim$ 0 and near phase
$\sim$ 0.8 at T $\sim$ 100 ks.}
\label{f:refine}
\end{center}
\end{figure*}

\section{Timing Analysis} \label{s:timing}

\subsection{Initial Search with Lomb-Scargle Periodograms} \label{s:initial}

We have searched for periodic X-ray modulations in the 296 sources
with $\ge$ 100 net counts in the 0.3--8 keV band from 31 individual
observations and 5 stacked data sets.  For each source, we applied barycenter
corrections on the arrival times of source and background events, and then
generated a background subtracted light curve in the 0.3--8 keV band with
a bin size of 3.14 s.  We created a Lomb-Scargle (LS) periodogram
\citep{Scargle82} from each light curve over a wide range of search periods.  

We determined independent search 
periods based on the overall observation duration and
the number of net counts. We followed the recipe by \citet{Horne86}, where
the number of the independent search frequencies ($N_f$) 
is given as --6.362+1.193 $N_E$ + 0.00098 $N_E^2$ and $N_E$ is the net
counts of the source. We selected $N_f$ periods from 2$T_D/N_E$ to
$T_D$ at an equal frequency spacing, where $T_D$ is the observation duration.
For example, a source with 1000 net counts observed in 100 ks
would require a search of 2167 independent periods from $\sim$ 100 s to 50 ks at a
frequency interval of $\sim$4.3 $\mu$Hz. In order to ensure a
successful search, we refine the frequency spacing further by a factor
of two (i.e.~$\sim$2.2 $\mu$Hz for the above example). 
In addition, we extend the search periods down to $\sim$6 sec, given the
large population of sub 100 sec period pulsars in the region. 
We keep the same frequency spacing when adding new periods. In the
above example, then, we would have scanned $\sim$87,000 periods and
roughly a half of them would be independent.

The false detection probability ($P_{FD}$) can be calculated from the
amplitude ($X$) of the LS periodogram and the total number of
search periods ($N_{T}$ = $\Sigma_i N_{f,i}$ where $N_{f,i}$ is the
the number of search periods for source $i$) as
$P_{FD}=1-(1-e^{-X})^{N_{T}}$.  We have searched about \nep{7}{6} independent
periods from the 296 sources in total, and corresponding 1\%, 5\% and
10\% false detection probabilities are at $X$~$\sim$~20.4, 18.7 and 18.0
respectively. To be conservative, we follow up on all the periods with
$X$~$>$~16.5 ($P_{FD}$ $\sim$ 50\%) for further analysis.  All of
the observations analyzed in this paper were conducted in the normal
\chandra ACIS readout mode where the CCD readout cycle time of
$\sim$~3.14 -- 3.24 s fundamentally limits the search for modulation periods $\lesssim$10~s.
Therefore, the above estimates of $P_{FD}$ are somewhat conservative.

Sources that are observed in a region with a noticeable exposure variation
($\gtrsim$10--20\%; e.g.,~near an edge of a CCD or a section with a large
CCD column-to-column efficiency variation) can exhibit false modulations
due to dithering motions. We excluded the known false periods (707 s
and 1000 s) and their harmonics from further investigation.

\subsection{Refining Pulsation Search with Epoch Folding} \label{s:refine}

To acquire precise modulation periods, we refined the initial
search results by performing the epoch folding (EF) search
\citep{Leahy83} and repeating the LS search for 1000 equally spaced
periods within the full-width half maximum (FWHM) of the candidate periods
from the initial search.  In the EF search, for a given search period,
we folded the source and background events in the 0.3--8 keV band in 15
equal phase bins.  The background-subtracted net counts in the folded
bins were converted to the folded light curve using the summed good time
intervals (GTIs) of each phase bin.  The reduced $\chi^2$ of the folded
light curve with respect to the constant rate was calculated to generate
periodograms of the EF searches.  

For a given initial candidate pulsation period, the refined LS and
EF periodograms present two new refined candidates, which do
not necessarily agree with each other.  To find the accurate modulation
period, we visually inspected the skewness of the X-ray event distribution
in the time versus phase diagrams at the two periods, and also
compared the modulation amplitudes
and pulsed fractions of the two
periods.  If one of the two periods
shows a noticeable skewness in the time versus phase diagram, the other
period is chosen to be the correct modulation period.  If neither of
the two periods shows any significant skewness, the period with a higher
modulation amplitude or a higher pulsed fraction (or both) is selected.
When the difference in the modulation amplitude and pulsed fraction
is marginal, we accept the result of the LS periodogram.

The modulation amplitude ($A$\Ss{mod}) is defined as
1--$r$\Ss{min}/$r$\Ss{max} where $r$\Ss{min} and $r$\Ss{max} are the
minimum and maximum of the folded light curve, respectively. The pulsed
fraction ($P_F$) is calculated as the ratio of the pulsating flux
above the minimum to the total flux, i.e., $\Sigma_i$
($r_i-r$\Ss{min})$/$$\Sigma_i$$r_i$, where $r_i$ is the count rate
of the folded bin $i$. In calculating $A$\Ss{mod} and $P_F$, we change the
number of the folded bins according to the total net counts in order to
limit the statistical fluctuation due to low count bins.  Each folded bin
should contain at least 25 net counts on average, but the total number of
the folded bins is no more than 20: e.g., 5, 10, 15, and 20 phase bins for
125, 250, 375, and $>$ 500 count sources, respectively.  In order to avoid
aliasing effects of binning, we generate multiple folded light
curves by varying the starting phase of folded bins and take the average
results of the multiple light curves for $A$\Ss{mod} and $P_F$.

Figure~\ref{f:refine} illustrates the selection process of the modulation
period for SXP~152 as an example.  Panel (a) compares the EF (black)
and LS (red) periodograms.  Panels (b) and (c) show the X-ray
event distribution in the elapsed observation time versus phase diagrams
folded at the peak periods of the EF and LS periodograms, respectively.
Two cycles are shown for easy viewing.  
In SXP~152,
the peak period of the LS periodogram (c) shows a noticeable drift
of the minimum phase (from phases~ $\sim$~0.7 at the early times to
phases~ $\sim$~0.8 at the late times), whereas the peak period of the EF
periodogram does not (b).  
Therefore, the pulsation period for SXP~152 is
determined to be 152.1063~$\pm$~0.0094~s.

The uncertainty of a modulation period is often estimated by an
1$\sigma$ equivalent spread of the peak in the periodogram. Instead we
follow the recipe given by \citet{Horne86}.  The error estimates by
the latter tend to be tighter than those by the former.  Our analysis
finds that the two peak periods of the refined LS and EF periodograms
are always consistent with each other within only a tiny fraction of the
uncertainty estimate by the former.  On the other hand, in some cases, the
visual inspection of skewness in the time versus phase diagram clearly
shows that one of the two periods has a noticeable skewness.  Therefore,
we conclude that the former approach is too conservative and the latter
produces more appropriate error estimates.
In the example shown in Figure~\ref{f:refine}, the two peak periods
of the EF and LS periodograms are within 
0.2$\sigma$ and 1.5$\sigma$ of each other under the peak width-based
error (0.058~s) and the uncertainty estimate (0.0094~s) by the
recipe in \citet{Horne86}, respectively. 

%%text_start

\subsection{Phase-Resolved Spectral Analysis} \label{s:spec}

We used energy-band dependent folded light curves, energy versus
phase diagrams, phase-segmented spectral fits and energy quantiles for
phased-resolved spectral analysis of the SMC pulsars.  For relatively
brightest sources with $\gtrsim$ 1000 net counts, we performed spectral
model fits of a few selected phase segments in order to constrain
the spectral parameters and their changes more precisely. Similarly
to the overall spectral model fits, except for the magnetar SXP8.02,
phase-resolved spectra of all the bright sources are better fitted by
an absorbed power-law model.

For fainter sources, X-ray color-color diagrams are often
used for the evaluation of two-parameter models for X-ray spectra with
poor statistics.  X-ray colors or hardness ratios, however, suffer a
spectral bias intrinsic to the sub-energy band selection. The Bayesian
Estimation of Hardness Ratio \citep[BEHR,][]{Park06} alleviates the
intrinsic bias to some degree through a rigorous probabilistic
treatment.  We use energy quantile diagrams consisting of median
energy and quartile ratio, which enable a bias-free evaluation of the
two-parameter spectral models \citep{Hong04}.

Equal-count phase bins were used for phase-resolved energy quantile
calculation in order to maintain roughly similar photon statistics
among different phase bins and
thus acquire a reliable estimate of energy quantiles for each phase bin.
For a given phase of a pulsation period, we first calculate the width of
each phase bin to include $\ge$ 50 net counts, and estimate the energy
quantiles of the phase bin accordingly. We repeat the calculation
for 100 different phase bins for every pulsar, where the number of the
independent bins is given as min[$\sim$$N_c$/50, 100] with $N_c$ 
the total net counts.
We explore whether the spectral variation, if any, is intrinsic
or absorption dependent through the phase-resolved quantile diagram.

We also use the energy versus phase diagrams to visualize diverse 
spectral changes over pulsation cycles.\footnote{For energy versus
phase diagrams, we ignore the background counts. This is acceptable
for bright SMC pulsars because of the relatively low background
in a small aperture of each source region, which is enabled by the superb
angular resolution of the \chandra X-ray optics.}
Folded-light curves in the three energy bands also
illustrate the spectral variation over pulsation cycles.

\begin{figure*} \begin{center}

\includegraphics*[width=0.988\textwidth,clip=true]{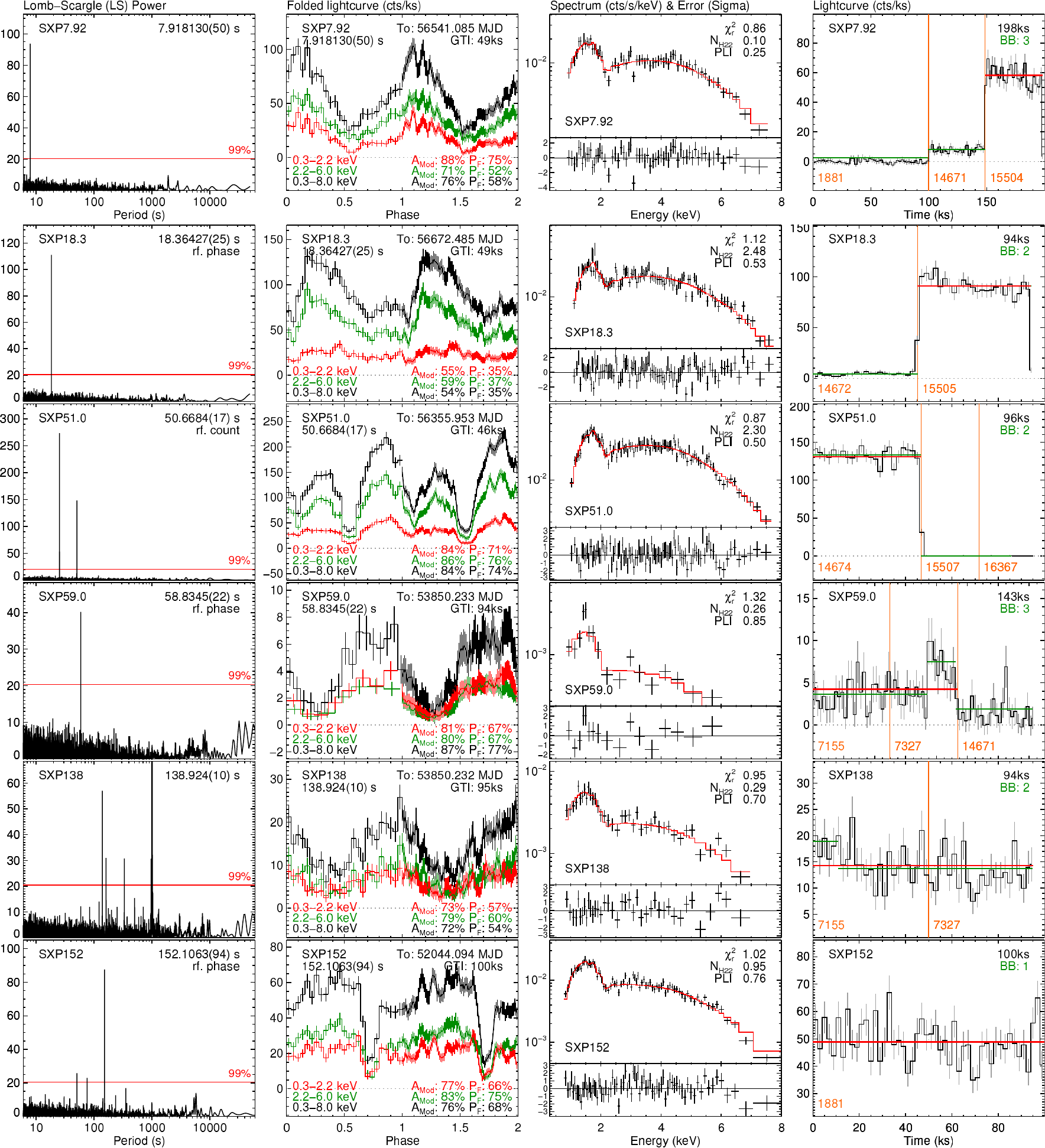}

\caption{LS periodograms, folded light curves, energy spectra, and compressed
light curves (i.e., without observational gaps) of the periodic X-ray
sources of SXP~7.92, SXP~18.3, SXP~51.0, SXP~59.0, SXP~138 and
SXP~152. The periodograms are shown
with the 99\% confidence levels (red horizontal lines,
Section~\ref{s:initial}). The labels 'rf.~phase' and 'rf.~count'
indicate that the pulsation period is selected from the refined EF
search instead of the LS periodogram (Section~\ref{s:refine}).
The folded light curves are drawn for the 0.3--8
keV (black), 0.3--2.2 keV (red), and 2.2--6 keV (green) bands along with
$A$\Ss{mod} and $P_F$.  In the energy spectra, 
the best-fit power-law models are shown in (red) lines.  In the compressed light
curves, the (red) horizontal
lines represent the average count rate of the observation(s) exhibiting the
periodicity and they indicate the observation(s) used for the folded
light curves and spectral fits (e.g., Obs.~IDs 7155 and 7327 for SXP~59.0), 
while the (green) steps show the
Bayesian Blocks (BBs) calculated from the unsubtracted event list
without observational gaps. 
The number of BBs are labeled on the
right-upper corner of each panel.  }
\label{f:pdA}
\end{center}
\end{figure*}

\begin{figure*} \begin{center}
\includegraphics*[width=0.988\textwidth,clip=true]{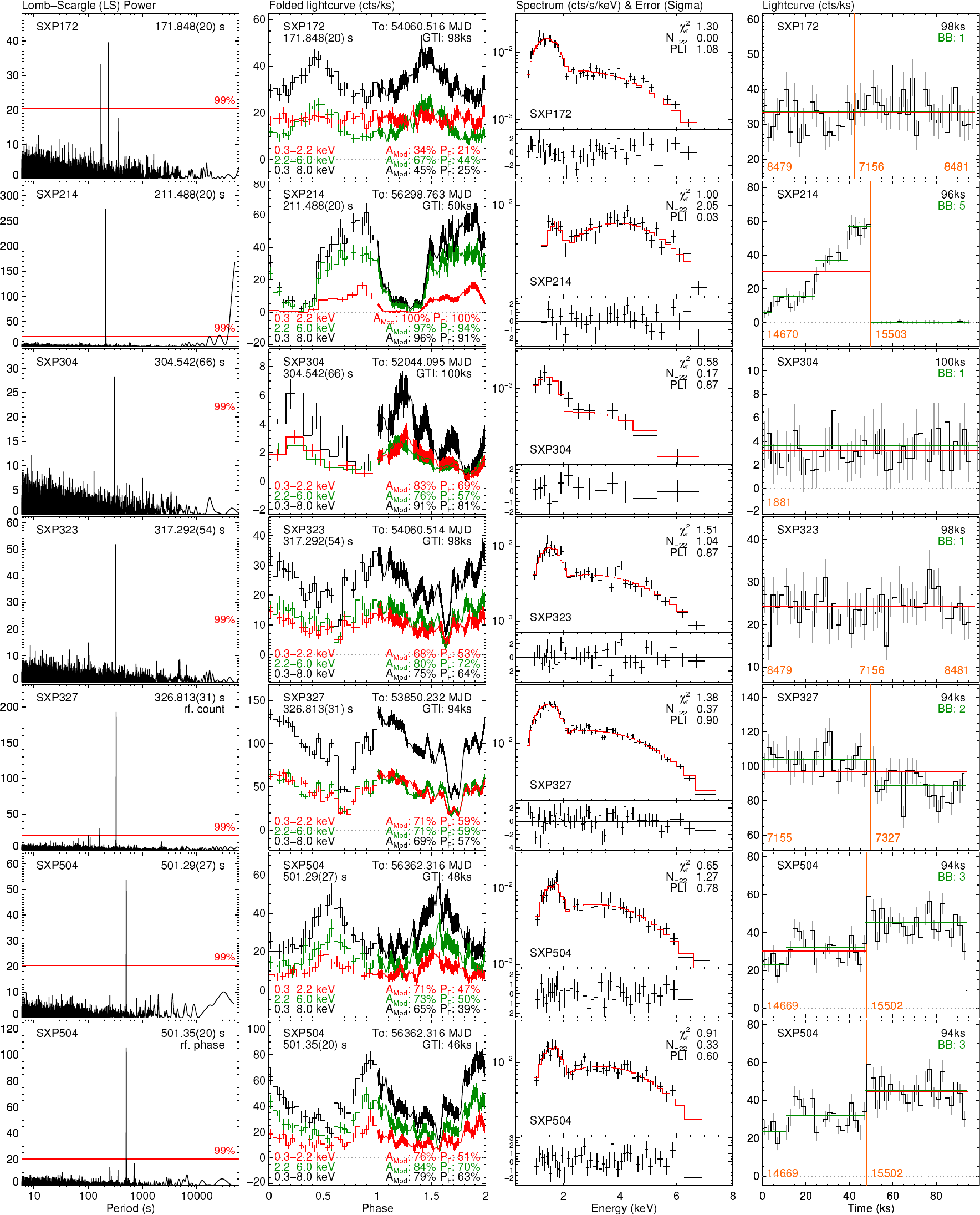}

\caption{Same as Figure~\ref{f:pdA} for SXP~172, SXP~214, SXP~304, SXP~323, SXP~327 and SXP~504. 
In SXP~504, the results of two observations (Obs.~IDs 14669 and 15502) are shown.
	}
\label{f:pdB}
\end{center}
\end{figure*}

\begin{figure*} \begin{center}
\includegraphics*[width=0.988\textwidth,clip=true]{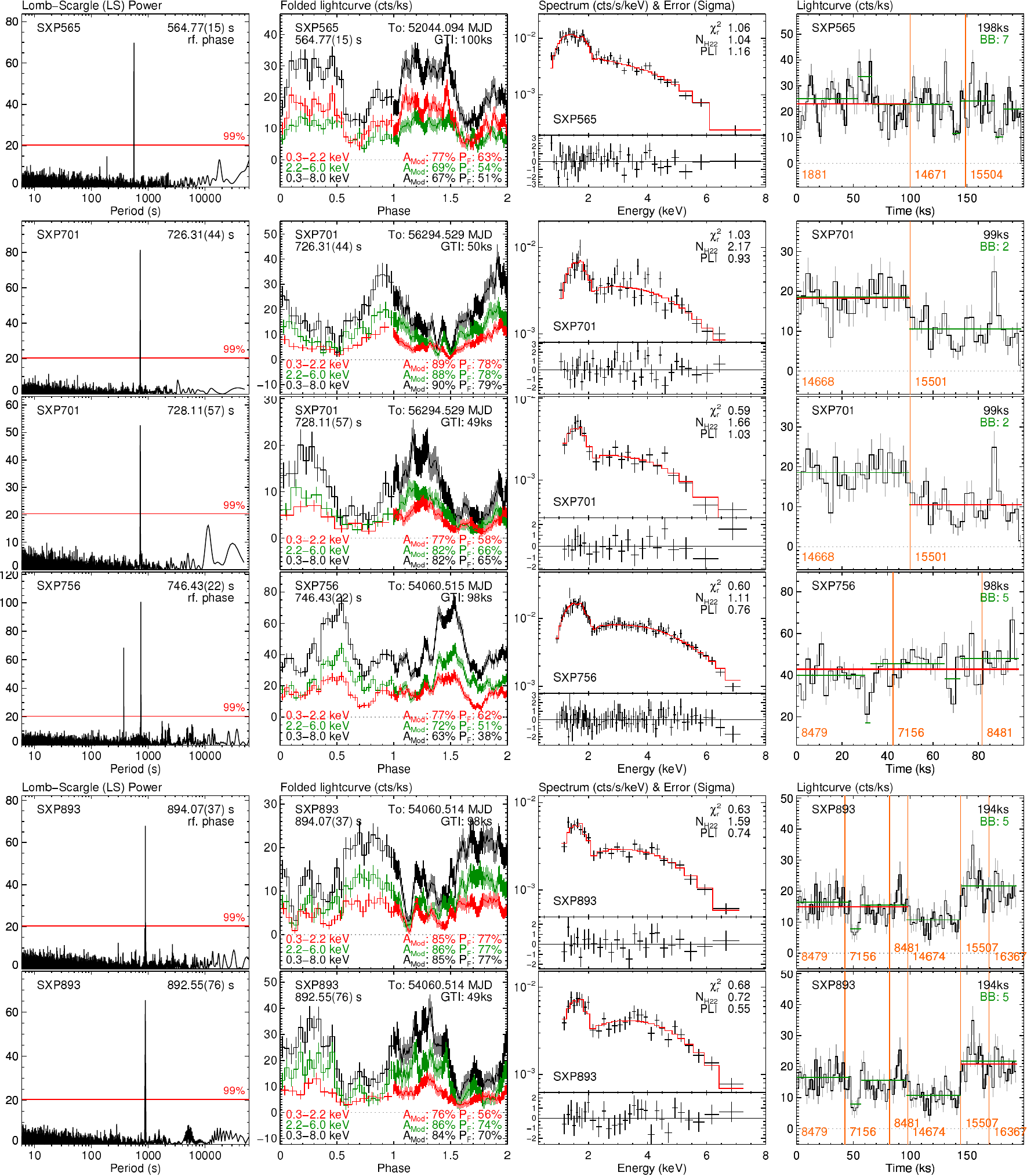}

\caption{Same as Figure~\ref{f:pdA} for SXP~565, SXP~701, SXP~756, and SXP~893.
In SXP~701 and SXP~893, the results of two sets of observations are shown: Obs.~IDs 14668 and 15501 for
SXP~701; Obs.~IDs 8478, 7156, 8481 and Obs.~IDs 15507, 16367 for SXP~893 (see Table~\ref{t:src}).
	}
\label{f:pdC}
\end{center}
\end{figure*}

\begin{figure*} \begin{center}
\includegraphics*[width=0.988\textwidth,clip=true]{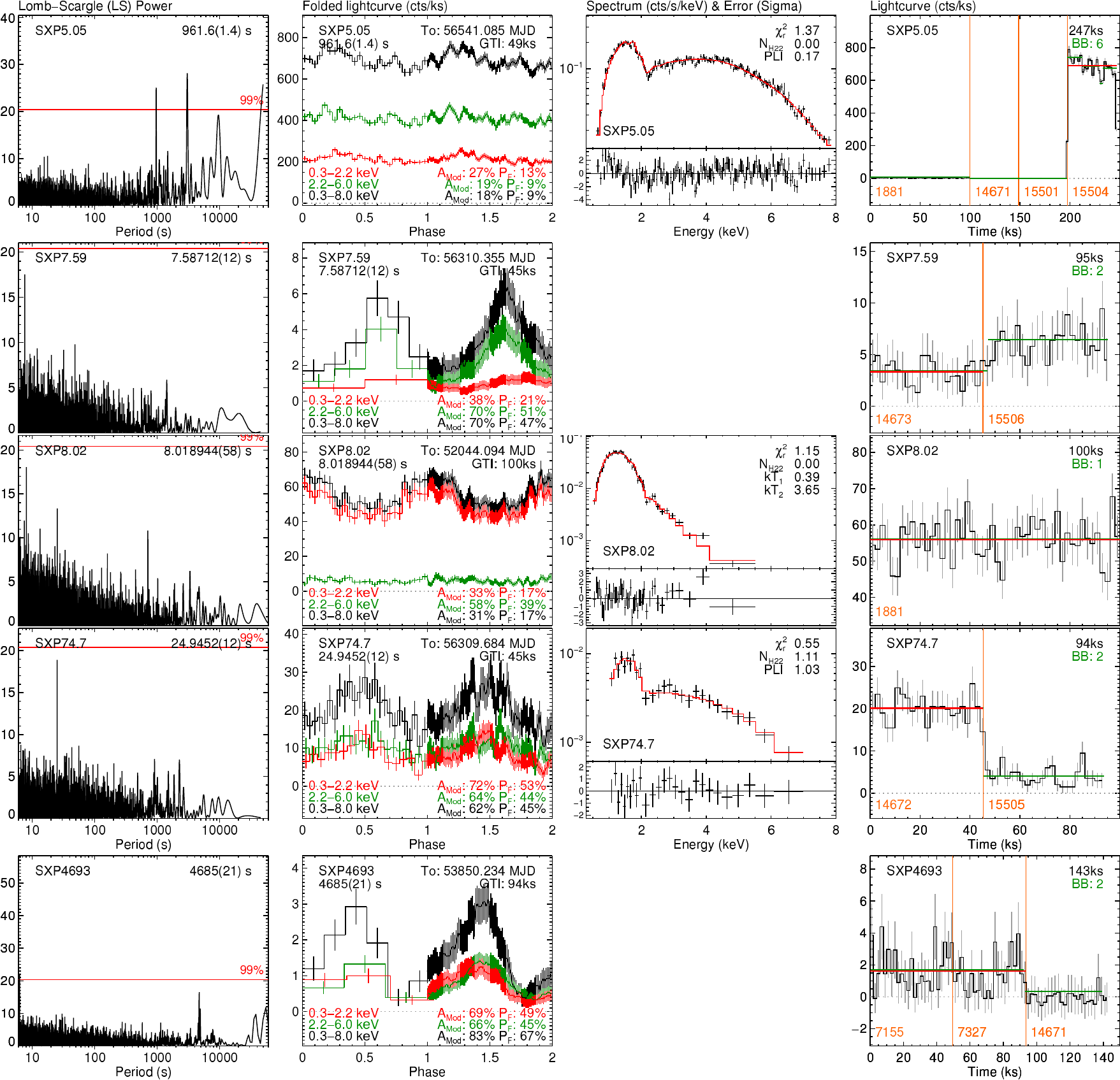}

\caption{Same as Figure~\ref{f:pdA} for SXP~5.05, SXP~7.59, SXP~8.02 and SXP~74.7 and SXP~4693. 
An absorbed two-temperature blackbody model is used for spectral model fit
for SXP~8.02. For spectral analysis of relatively low count sources
(SXP~7.59 and SXP~4693, $\sim$150 net counts), we rely on quantile analysis 
instead of spectral model fitting (Table~\ref{t:spec}).}
\label{f:pdD}
\end{center}
\end{figure*}

\section{Pulsation Search Results} \label{s:pulsars}

Table~\ref{t:src} summarizes our pulsation search results,
which are grouped into three categories: (candidate) pulsars
exhibiting solid, marginal, and no pulsations during our survey.
All the periodic modulations are from the known pulsars except for
a relatively marginal detection of periodicity at 7.59~s from CXOU
J003942.37--732427.4. SXP~51.0, SXP~214 and SXP~701
show notable deviations in the periods from the
reported values in the literature \citep[see below and ][]{Hong16}.
SXP~25.5 and SXP~51.0 are identified as the same source with the
latter representing the proper spin period.

For the pulsars exhibiting no significant periodic modulations in the
LS periodograms, we manually performed the EF searches and calculate the
upper limit of $A$\Ss{mod} around the known pulsation periods.  For the
known pulsars that were not detected in our source search, we estimated
the upper limit of the flux and luminosity based on the background counts
in the 95\% point spread function (PSF) around the source position.
Flags in Table~\ref{t:src} indicate sources with the following possible
spectral variations based on phase-resolved quantile diagrams: "i" for
intrinsic variation (i.e., changes in $\Gamma$ for an absorbed
power-law model) and "a"  for changes more likely in absorption (\nH).

Among the pulsars without no pulsations observed with \chandra, 
the identification of SXP~9.13 as the \rosat source RX J0049.5--7311
is in question due to lack of pulsations from the source even though
the source was observed with \xmm and \chandra many times
\citep[see][]{Haberl16}. Its pulsation was originally discovered from
the \asca source AX J0049-732.  According to the source poisition in
the catalogue
by \citet{Coe15}, SXP~280 should have been covered by the \chandra
observation (Obs~ID. 1881), but 
\citet{Haberl16} showed that its optical countpart
is located outside of the \chandra field of view.
Therefore, SXP~280 is excluded in the list.

Figures~\ref{f:pdA}--\ref{f:pdD} show the LS periodograms, the folded
light curves, the overall spectra with model fits, and the compressed
light curves (without observational gaps) of the 21 SMC pulsars exhibiting
pulsations during our survey.  
The LS periodograms cover $\sim$~5~s to $\sim$ 50 ks, and the red
horizontal line indicates a confidence level of 99\% ($P_{FD}$ =
1\%).  The folded light curves of the equal phase bins are shown in
the three energy bands (black: 0.3--8 keV, red: 0.3--2.2 keV, green:
2.2--6 keV) for two cycles: the first cycle (phase 0 to 1) is shown
with independent phase bins, whereas the second cycle (phase 1 to 2)
is shown in a sliding window of phase bins. 

For spectral model fits, we
have tried an absorbed power-law, thermal bremsstrahlung, blackbody, and APEC
models with two absorption components: the Galactic foreground absorption
of \nH=\nep{6}{20} cm\sS{-2} \citep{Dickey90} with solar abundance 
($Z$~=~$\Zs$), and the free SMC local absorption with
$Z$~=~0.2~$\Zs$ \citep{Russell92} using the absorption model by
\citet{Wilms00}.  All except for SXP~8.02 are best fitted by an
absorbed power-law model with $\Gamma$ $\lesssim$
1.5 among the four models considered here.  SXP~8.02, a magnetar, is better fitted
with a two-temperature blackbody model (Section \ref{s:sxp8.02}).
Table ~\ref{t:spec} summarizes the spectral properties
of the 31 SMC pulsars covered in the survey, which also includes
all the results from the observations even when no pulsations
were detected.

The compressed light curves show the changes in the background-subtracted
event rates without observational gaps. The average rate
during the observation(s) exhibiting the pulsation is shown in red, and
the green segments show the average rate of the independent Bayesian Blocks (BBs), 
which mark the change points between time intervals of statistically
different rates \citep{Scargle13}.

%%text_stop

\begin{figure*} 
\begin{center}

\includegraphics*[width=0.90\textwidth,clip=true]{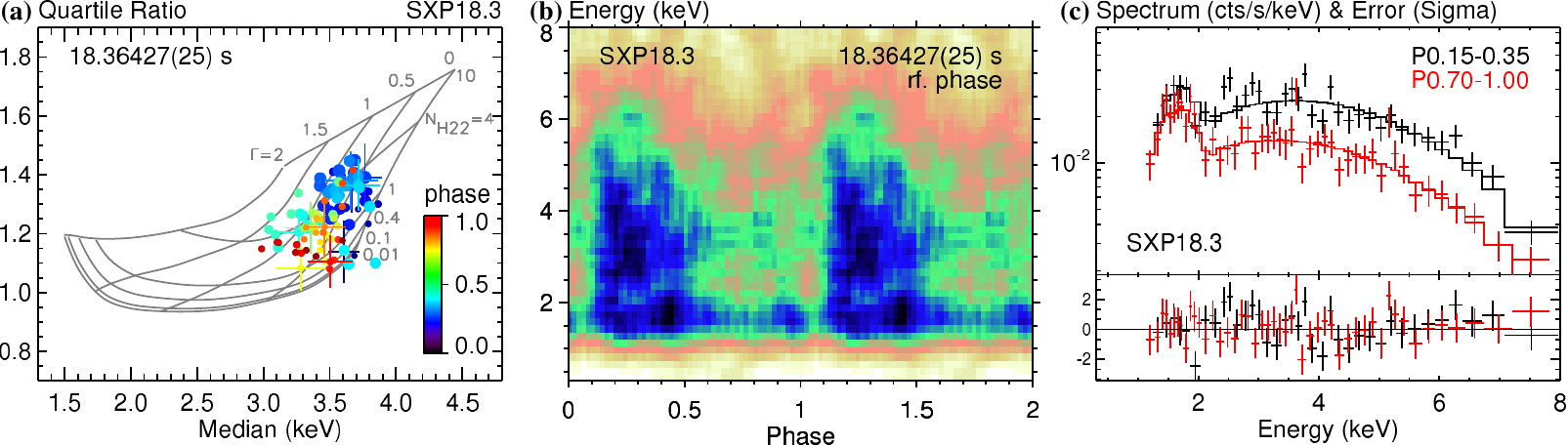}
\caption{
Phase-resolved quantile diagram, energy versus phase diagram
and spectral model fits of SXP~18.3.  See Section~\ref{s:SXP18.3}.
The data points in the quantile diagrams are color-coded by pulsation
phases (i.e., purple at phase~$\sim$~0.0 and red at phase~$\sim$~1.0),
and the grids are for absorbed power-law models with $\Gamma$ = 0, 0.5,
1, 1.5, and 2, and \nH = 0.01, 0.1, 0.4, 1, 4, and \nep{10}{22}
cm\sS{-2}. 
}
\label{f:sxp18.3}
\end{center}
\end{figure*}

Below  we explore the X-ray properties of
each SXP that exhibited X-ray modulations during the \chandra
observations. X-ray luminosities quoted in the following sections
refer to the observed values in the 0.5--8 keV band,
assuming a distance of 60 kpc unless specified
otherwise. Table~\ref{t:spec} lists both the observed and intrinsic
X-ray luminosities.

\begin{figure*} 
\begin{center}

\includegraphics*[width=0.90\textwidth,clip=true]{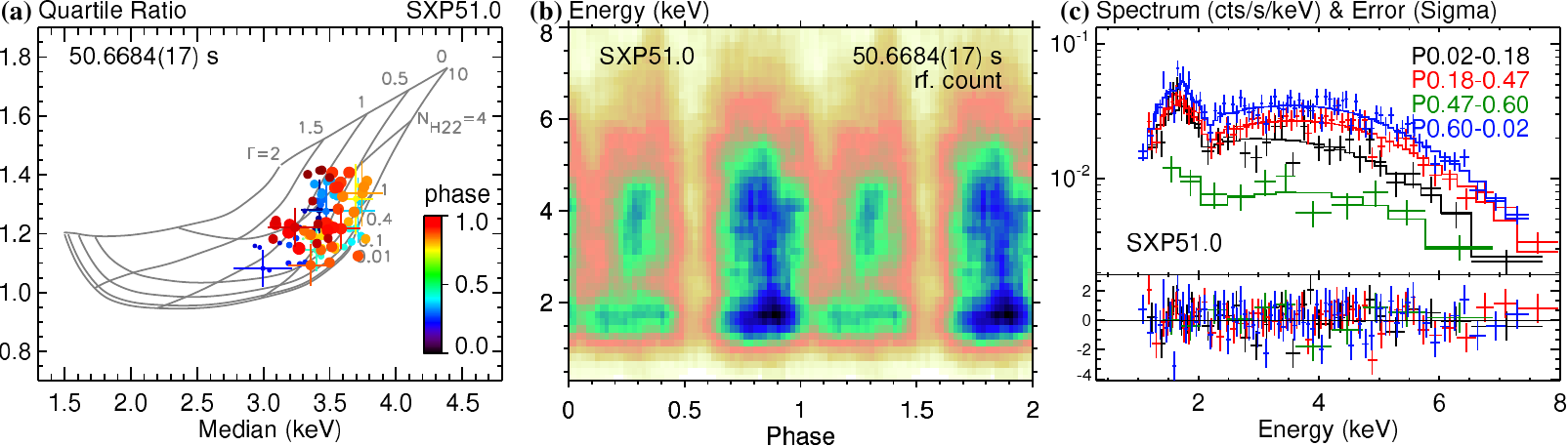}
\caption{
Phase-resolved quantile diagram, energy versus phase diagram
and spectral model fits of SXP~51.0.  See Section~\ref{s:SXP51.0}.
}
\label{f:sxp51.0}
\end{center}
\end{figure*}

%%text_stop

%%text_start
\subsection{SXP~7.92} 

SXP~7.92 was first identified as a pulsar from 
an \rxte observation by \citet{Corbet08}.
\citet{Coe09} suggested the early-type star AzV285 with a spectral type
of O9-B0 III as the counterpart. 

SXP~7.92 was observed with \chandra once in 2001 and twice in 2013. In
2001, the source was positioned in ACIS-S3, far off the aimpoint, and
it was not detected with an upper limit (3$\sigma$) of the 0.5--8 keV
X-ray luminosity at 2.4$\times$10\sS{33} \lcgs for an absorbed power-law
model (see below for the model parameters).
We detected the source with an X-ray luminosity of
$\sim$10\sS{35}~\lcgs in the 2013 January observation, but no
pulsation was observed ($A$\Ss{mod} $\lesssim$ 46\%).
The observation eight months later showed that the X-ray luminosity had
increased to 7.5$\times$10\sS{35} \lcgs,
and a significant pulsation was
detected. The LS periodogram shows strong peaks at 5.5088 and 7.9181 s,
and the former is the result of beating between the pulsation period
(the latter) and the CCD readout cycle ($\sim$~3.14~s).
The observed luminosity is still lower than what \swift observed in 2008
by an order of magnitude \citep{Coe09}, indicating the large flux changes
over four orders of magnitudes.

\begin{figure*} 
\begin{center}

\includegraphics*[width=0.90\textwidth,clip=true]{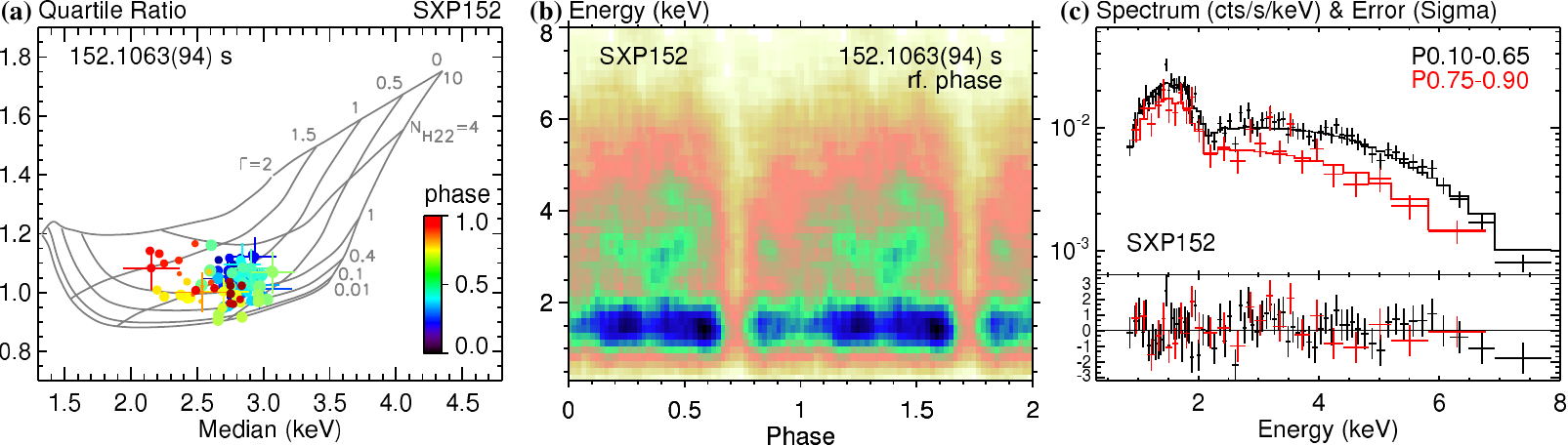}
\caption{Phase-resolved quantile diagram, energy versus phase diagram
and spectral model fits of SXP~152.  See Section~\ref{s:SXP152}.
}
\label{f:sxp152}
\end{center}
\end{figure*}

The overall X-ray spectrum (Obs.~ID 15504) is well fitted by an absorbed
power-law model with $\Gamma$~=~0.25 $\pm$ 0.06 and
\nH~=~10\sS{21} cm\sS{-2}.  The relatively hard \chandra spectrum is consistent
with the quantile analysis from the 2008 \swift data
by \citet{Coe09}.  The X-ray spectrum did not show any significant change
between the two observations in 2013.  Despite the relatively short
period that can cause the phase mixing for the given CCD readout cycle,
the pulse profile shows a large modulation
amplitude ($A$\Ss{mod}~$\sim$~76\%).  The pulse profile of SXP~7.92 is
known to transition between the single peak and two peak shapes \citep{Coe09}. The
pulse profile measured with \chandra shows a single peak similar to what
the 2004 \rxte observation detected \citep{Coe09}, but, given the
short pulsation period, the long CCD readout cycle in the \chandra
ACIS chips may have contributed to the apparent single peak
profile. The phase-resolved quantile
diagram shows a marginal variation in the X-ray absorption but no apparent
correlation with the flux.

\begin{figure*} 
\begin{center}
\includegraphics*[width=0.90\textwidth,clip=true]{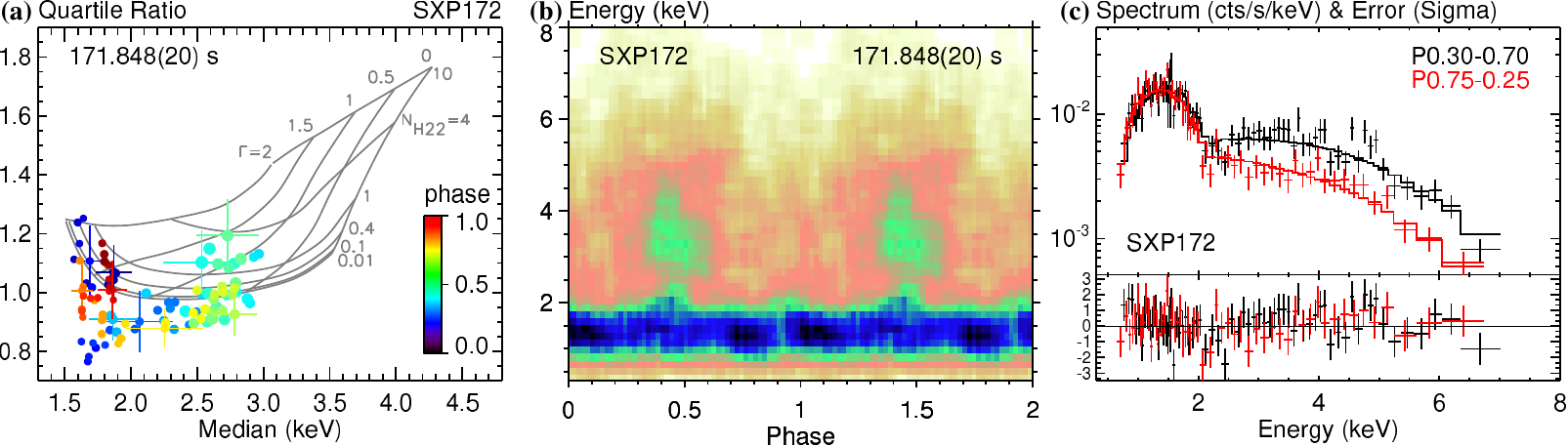}
\caption{Phase-resolved quantile diagram, energy versus phase diagram
and spectral model fits of SXP~172.
See Section~\ref{s:SXP172}.
}
\label{f:sxp172}
\end{center}
\end{figure*}

\subsection{SXP~18.3} \label{s:SXP18.3}

X-ray pulsations from SXP~18.3 were first found with \rxte
in 2003 by \citet{Corbet03}. The position of the source was
later localized with \xmm by \citet{Eger08}. The OGLE-III optical
light curve revealed a coherent period of $\sim$ 18 d \citep{Udalski08}. 

The source
was observed with \chandra in January of 2014 and 2015. Between the two
observations, the X-ray luminosity increased from \nep{3.8}{34}
to 10\sS{36} \lcgs. The latter is still lower than what was
observed during the 2007 \xmm observation \citep{Eger08} or the average X-ray
luminosity seen with \rxte \citep{Klus14} by a factor of $\sim$6--7. 
The X-ray pulsation is not detected in the 2014 observation
($A$\Ss{mod} $\lesssim$ 66\%), and the pulsation period detected in the 2015
observation ($A$\Ss{mod} $\sim$ 54\%) is consistent with the long-term 
spin-up trend reported by \citet{Klus14}. 

The X-ray spectrum, well fitted by
an absorbed power-law model with $\Gamma$ = 0.53 $\pm$ 0.06 (Figure~\ref{f:pdA}), 
remained consistent between the \xmm and two \chandra observations. 
The phase-resolved quantile diagram in
Figure~\ref{f:sxp18.3}(a) shows a marginal correlation between the
absorption and the flux, where 
a heavier absorption is present during phases of higher fluxes.
The spectral fits of the two selected phases representing the peak
(phases 0.15--0.35) and
valley (phases 0.7-1.0) of the folded light curve show similar results 
(Figure~\ref{f:sxp18.3}(b)). Table~\ref{t:phrs_sxp18.3} summarizes
the best-fit spectral parameters.

\subsection{SXP~51.0} \label{s:SXP51.0}

X-ray pulsations at 25.5 s from SXP~51.0 were first discovered with \rxte
by \citet{Lamb02a}. The source was later localized as XMMU
J004814.1--731003 from an \xmm observation by \citet{Haberl08} and has
been known as SXP~25.5 for some time.  It has been unclear if XMMU
J004814.1--731003 is associated with AX J0048.2--7309, which is listed
as SXP~51.0 in the literature.  \citet{Laycock05} showed that the
1/2nd harmonic of a period can dominate the power spectrum and claimed
that the 25.5~s period is a harmonic of the 51 s period, whereas
\citet{Haberl08} failed to detect the 51 s period from the \xmm observation.  

SXP~51.0 was observed with \chandra in 2013 March and September. In
the March observation, the source was bright with an X-ray luminosity
of 1.7$\times$10\sS{36} \lcgs, which is consistent with the average
luminosity observed with \rxte \citep{Klus14} within a factor of two,
but in September the X-ray luminosity dropped below the 3$\sigma$
detection limit of 1.4$\times$10\sS{33} \lcgs.

The elongated event distribution of the source seen in the 2013 March
\chandra observation may suggest a possibility of two sources, but a
large offset (5.5\arcmin) from the aimpoint is likely the origin of
the wide PSF.  The LS periodogram from the 2013 March \chandra
observation shows strong pulsations at both 25.33426 $\pm$ 0.00032 s
and 50.6684 $\pm$ 0.0017~s periods with $\sim$25.3~s as the primary
periodicity.  Given the simultaneous detection of the pulsations at
both periods, we believe that SXP~25.5 and SXP~51.0 are the same source.

\citet{Arumugasamy14} argued that the spin period of PSR J2022$+$3842
is 48.6 ms even though they found a stronger signal at 24.3 ms in their
periodogram because the folded light curve at 48.6 ms showed two distinct
peaks per pulsation cycle.
Similarly \citet{Mori13} claimed the pulsation period of 3.76 s for
SGR J1745--29 even though their power spectrum shows a stronger peak at 1.25 s. 
If we follow the same logic, the true spin period of the source is 50.6684~s
since its pulse profile 
shows two distinct peaks with two sharp eclipse-like dips
(Figure~\ref{f:pdA}). 
On the other hand, the pulse profile folded at
the primary period of 25.33426~s shows a somewhat featureless
single broad peak.
SXP~51.0 being misidentified as SXP~25.5 is somewhat intriguing since
the apparent bimodal distribution of the spin periods among the
pulsars pointed out by \citet{Knigge11} is due to a
paucity of pulsars with spin periods around 25 s. 

The observed primary period at 25.33426~s, in fact, is
significantly shorter than the spin periods reported from the previous
\xmm observations \citep[e.g., $\gtrsim$25.45~s in][]{Klus14}.
Figure~B10 in \citet{Klus14} shows that the source is recently on a
long-term spin-up trend of $\dot{P}$ $\sim$ --0.10 s yr\sS{-1} from MJD
55200 to 55700 (assuming $P$ $\sim$ 51~s instead of $\sim$ 25.5~s). The
spin-up trend between our measurement (MJD
56355) and the last \xmm observation is $\dot{P}$ $\sim$ --0.20 s
yr\sS{-1}, which indicates that the spin-up trend may have been accelerated.

The X-ray spectrum is well fitted by an absorbed power-law model with
$\Gamma$ = 0.50 $\pm$ 0.05, which is relatively harder
than $\Gamma$ = 1.3 $\pm$ 0.3 measured by \citet{Haberl08}, whereas the
absorption in the spectrum remained consistent in both measurements.
The phase-resolved quantile diagram shows spectral variations
with the pulsation phase as shown in Figure~\ref{f:sxp51.0}(a) \& (b). 
The two peaks in the pulse profile exhibit a
consistent spectral type with a marginal line absorption feature
around 5 keV ($\sim$ 3$\sigma$), 
whereas the gaps between the peaks show distinct spectral
types of either being intrinsically softer or less absorbed.
See Figure~\ref{f:sxp51.0}(c) and Table~\ref{t:phrs_sxp51.0}.

\subsection{SXP~59.0}
SXP~59.0 was first discovered as a transient pulsar in 1998 with \rxte
by \citet{Marshall98}. \citet{Laycock05} derived an orbital period of
123 $\pm$ 1 d from four bright X-ray bursts seen with \rxte in 1998
and 1999. \citet{Schmidtke05} proposed an orbital period of 60.2 d
from a timing analysis of the OGLE-II and MACHO data. However,
\citet{Galache08} suggested an orbital period of 122.1 $\pm$
0.38 d based on an analysis of the \rxte data between 1999 September and
2002 September.

SXP~59.0 was observed with \chandra twice in 2006 and once in 2013.
Between the 2006 and 2013 observations
the X-ray luminosity decreased to 10\sS{34} \lcgs from \nep{4}{34} \lcgs
and the X-ray spectrum became intrinsically softer ($\Gamma$ $\sim$ 2.6
from 1) but more absorbed (\nH $\sim$ \nep{6}{22} cm\sS{-2} from 10\sS{21} 
cm\sS{-2}) according to the quantile analysis.  
The X-ray spectrum observed in 2006 is best fitted by an
absorbed power-law model with $\Gamma$ = 0.9 $\pm$
0.2 among the four basic models but the reduced $\chi^2$ is relatively high
(1.32).  The 2013 observation detected only 67 net counts from the
source (Table~\ref{t:spec}).

The periodic modulation is detected in the former data set
($A$\Ss{mod}~$\sim$~80\%), but in the latter the low statistics 
limits the modulation search.  The 2006
light curve shows a sudden increase in the flux (Figure~\ref{f:pdA}),
and the BB analysis on the compressed light curve shows
three distinct blocks with the second block marking the increase in the
flux. 

\subsection{SXP~138}
Pulsations from SXP~138 were first discovered in the
\chandra archival data by \citet{Edge04b}.
\citet{Edge05} identified  MA[93]667 as the optical counterpart
and found an orbital period of $\sim$125.1 d from the MACHO data.
SXP~138 was observed again with \chandra in 2006.
The LS periodogram in Figure~\ref{f:pdA} shows significant
periodicities ($A$\Ss{mod} $\sim$
72\%) at the spin period (138.924 $\pm$ 0.010 s) as well as the
observational dithering period (1000 s) and their beating periods
since the source was detected near a CCD edge.
The overall light curve shows a marginal decline with an average X-ray
luminosity of $\sim$\nep{2.4}{35} \lcgs (2 BBs in  Figure~\ref{f:pdA}). 
The overall X-ray
spectrum is relatively hard with $\Gamma$ = 0.70 $\pm$ 0.09 for an absorbed
power-law model.

\subsection{SXP~152} \label{s:SXP152}
Pulsations from SXP~152 were first discovered
from a \chandra observation in 2001 by \citet{Macomb03}.
Our analysis covers the same data set.
The LS periodogram from our analysis shows the main
pulsation at 152.1063 $\pm$ 0.0094~s.
The 1/2 (76.04 s) and 1/3 (50.70 s) harmonics of
the main period all form significant peaks in the LS periodogram.

The folded light curve at the main period shows a sharp dip or an
eclipse-like feature spanning over $\sim$0.15 in phase
($A$\Ss{mod}=76\%). This feature was used to improve the accuracy in
measurement of the pulsation period (Section~\ref{s:refine} and
Figure~\ref{f:refine}).
For instance, the pulsation period found in our analysis is consistent with 152.098
$\pm$ 0.016 s reported by \citet{Macomb03}.  The event distribution in
the time versus phase diagram at the latter period, however, shows
that the dip in the pulse profile slips in phase with the
time, similarly to the case in Figure~\ref{f:refine}(c).  
Therefore, we believe that our estimate of the pulsation period is
more precise.

The overall spectrum is well fitted by
an absorbed power-law model with $\Gamma$ = 0.76 $\pm$
0.04. Our estimate of \nH = \nep{9}{21} cm\sS{-2} is higher than
\nep{5.7}{21} cm\sS{-2} by \citet{Macomb03}, which could be
due to the different assumption of the absorption models.
The observed 0.5--8 keV X-ray luminosity (\nep{5.2}{35} \lcgs)
is about a factor of two higher than what
\citet{Macomb03} estimated, which can be explained by the differences in the
energy band (0.6--7.5 keV in the latter), the assumption for the
distance to the source (57 kpc in the latter), and spectral model parameters.

The phase-resolved quantile diagram shows that the X-ray spectrum
during the plateau right after the dip is intrinsically softer than
the rest for an absorbed power-law model (a few red points are at
$\Gamma$ $>$ 1 in the quantile diagram in Figure~\ref{f:sxp152}(a),
see also Figure~\ref{f:pdA}).  Figure~\ref{f:sxp152}(c) and
Table~\ref{t:phrs_sxp152} compare the spectral model fit results of
two selected phases, where a marginal spectral softening is observed
after the dip.  The folded light curve shows rapid fluctuations up to the
ingress of the dip.  Similarly to SXP~51.0, when the flux is high
(phases 0.1--0.65), the X-ray spectrum of SXP~152 exhibits marginal
($\sim$3$\sigma$) absorption features near 3.5 keV and 5 keV
(Figure~\ref{f:sxp152}(c)).  The absorption feature near 3.5--4 keV is
identifiable in the energy versus phase diagram
(Figure~\ref{f:sxp152}(b)) from a pocket of the lower count region
near phases 0.2--0.5.

\subsection{SXP~172} \label{s:SXP172}

The X-ray pulsation from SXP~172 was first discovered by 
\citet{Yokogawa00a} from three \asca observations of the source in
1997, 1999, and 2000.
\citet{Haberl04} confirmed the X-ray pulsations at 172.21 $\pm$ 0.13 s
from an \xmm observation in 2000.  SXP~172 was observed with \chandra
in 2006. The LS periodogram shows two significant periodicities at
171.848 $\pm$ 0.020 s and 235.831 $\pm$ 0.034 s with the latter being
more significant. 
The source was observed near a corner of a CCD during
the \chandra observation, but the usual dithering periods at 707 s and
1000 s are not present in the LS periodogram. 
The newly detected period ($\sim$ 235 s) appears to be an artifact
due to a filtering between Level 1 and 2 standard CXC data products,
which removed some events in the central section
of the PSF. Only the Level 2 data exhibits the pulsations at $\sim$ 235 s while
the pulsations at $\sim$ 172 s are present in both the Levels 1 and 2 data.

With an average X-ray luminosity of \nep{3.6}{35} \lcgs, the overall X-ray
spectrum is relatively hard with $\Gamma$ = 1.08 $\pm$ 0.03
for an absorbed power-law model.
At the main 172 s period, the phase-resolved quantile diagram
indicates that the high flux is correlated with an intrinsic spectral
hardening (lower photon indices).  Figure~\ref{f:sxp172}(a and c) and Table~\ref{t:phrs_sxp172} show that
$\Gamma$ switches between $<$1 and $>$1 during the pulsation.  
The energy versus phase diagram in Figure~\ref{f:sxp172}(b) and the
multiband folded light
curves in Figure~\ref{f:pdB} show that the modulation is dominated by
the hard X-ray photons above 2 keV, where the hard X-ray emission
peaks in phases between 0.3 and 0.7. 

\subsection{SXP~214} 
SXP~214 was discovered as a transient pulsar with a
spin period of $P_s$ = 214 s from an \xmm observation
in 2009 \citep{Coe11}.
The \chandra observation of SXP~214 in 2013 shows interesting spectral
variations as a function of time (Figure~\ref{f:pdB}) and the
pulsation phase. \citet{Hong16} have concluded that the 
NS was emerging from the circumstellar disk of
the Be companion star during the \chandra observation.
See \citet{Hong16} for the detailed analysis and results of the \chandra observation
of SXP~214.

\begin{figure} 
\begin{center}
\includegraphics*[width=0.30\textwidth,clip=true, trim=0.63cm 0.0cm 0.3cm 0.1cm]
{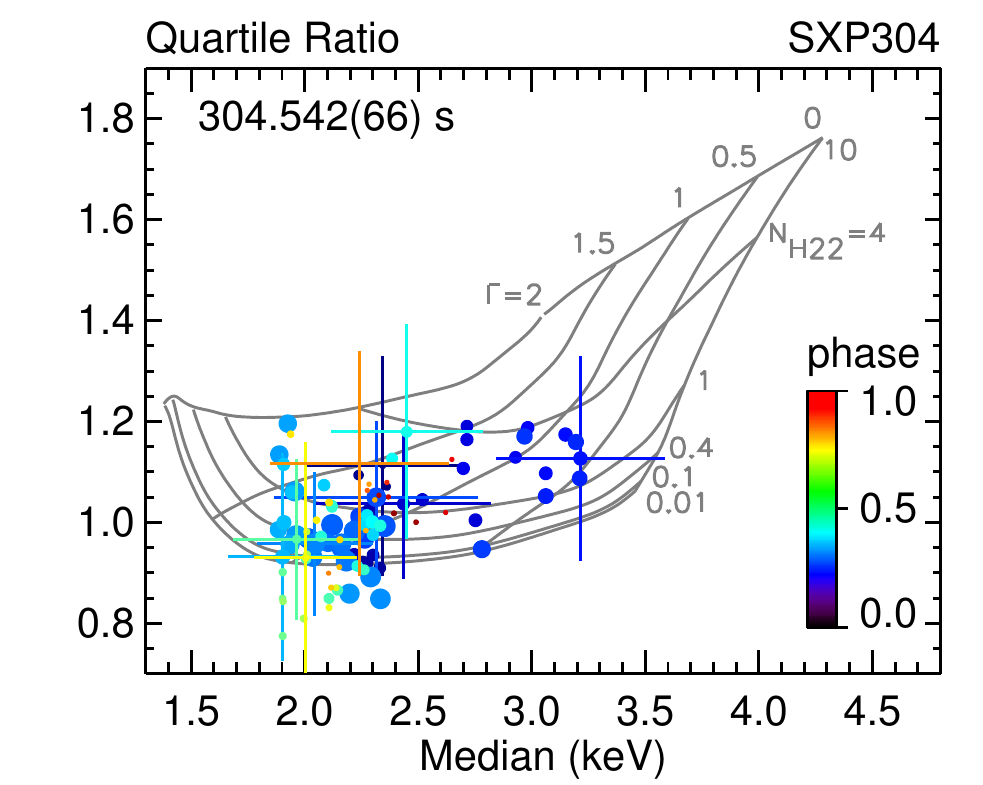}
\caption{Phase-resolved quantile diagram of SXP~304. See Section~\ref{s:SXP304}.}
\label{f:sxp304}
\end{center}
\end{figure}

\begin{figure*} 
\begin{center}

\includegraphics*[width=0.90\textwidth,clip=true]{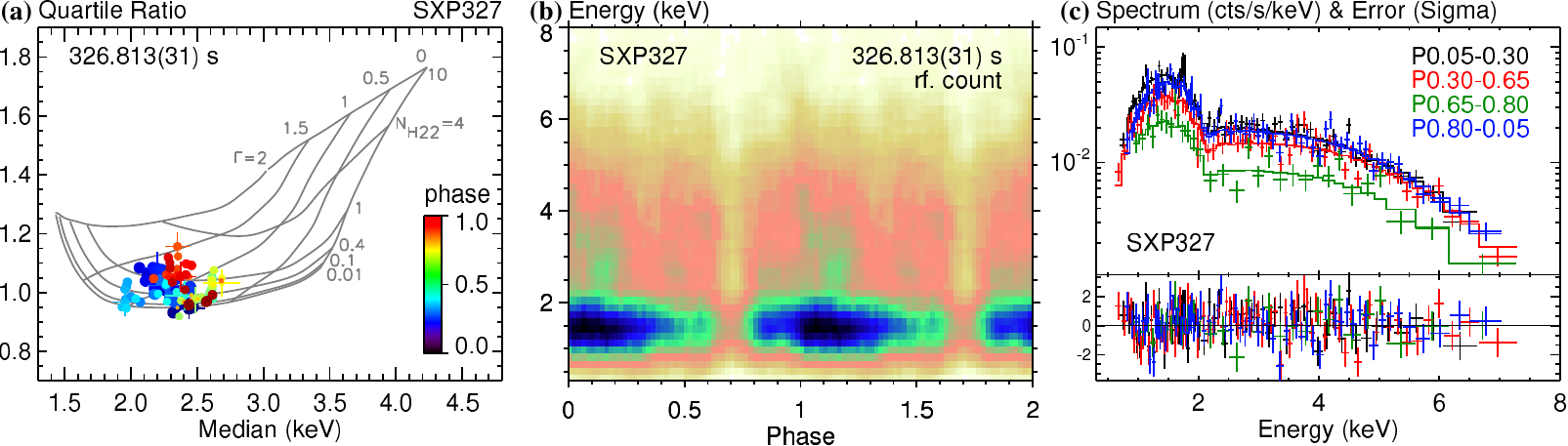}
\caption{Phase-resolved quantile diagram, energy versus phase diagram
and spectral model fits of SXP~327. See Section~\ref{s:SXP327}.} 
\label{f:sxp327}
\end{center}
\end{figure*}

\subsection{SXP~304} \label{s:SXP304}
Pulsations from SXP~304 were first discovered from
a 2001 \chandra observation by \citet{Macomb03}. Our analysis covers the same data set.
\citet{Schmidtke06} estimated an orbital period of 520 $\pm$
12 d from the MACHO data of the optical counterpart MA[93]1240.

The observed pulsation period (304.542 $\pm$ 0.066 s) is
consistent with the reported value by \citet{Macomb03}.
The pulse profile shows a high modulation amplitude
($A$\Ss{mod} $>$ 80\%) and the X-ray luminosity is the lowest
($\sim$\nep{3}{34} \lcgs in 0.5--8 keV) among the sources exhibiting
solid periodic
modulations.  Similarly to SXP~152, our spectral model fit of SXP~304
estimates
a lower extinction (\nH = \nep{2}{21} cm\sS{-2}) 
and a lower photon index ($\Gamma$ = 0.9 $\pm$ 02)
than what \citet{Macomb03} estimated. 
The differences in the model parameters
likely contributed to our higher estimate of the X-ray
luminosity (by a factor of $\sim$3).
The phase-resolved quantile diagram shows a marginal
spectral softening ($\Gamma$ $>$ 1) associated with the flux decrease during the
pulsation cycle (Figure~\ref{f:sxp304}).

\subsection{SXP~323}
Pulsations from SXP~323 were first discovered from an \asca observation
by \citet{Yokogawa98}. \citet{Cowley97} identified
a Be star as the optical counterpart. \citet{Laycock05}
suggested an orbital period of 109 $\pm$ 18 days from an early part of
the \rxte data, and \citet{Galache08} suggested an orbital period
of 116.6 $\pm$ 0.6 days from the 9 yr \rxte data excluding
the outburst at circa MJD 52,960. 

SXP~323 was observed with \chandra three times in 2006. The observed pulsation
period (317.292 $\pm$ 0.054 s) is consistent with the overall
spin-up trend seen in the \rxte observations \citep{Klus14}.  The
pulse profile shows an eclipse-like sharp dip in phases between 0.6
and 0.7 (Figure~\ref{f:pdB}). The overall X-ray spectrum is marginally fitted
($\chi^2_r$ $\sim$ 1.5) by an absorbed power-law model with 
$\Gamma$ = 0.87 $\pm$ 0.07, which indicates that an additional
component may be needed: the residual errors in the spectral fit show possible line
absorption and emission features in the 3.5 to 4 keV band (Figure~\ref{f:pdB}). 
The observed X-ray luminosity was about \nep{2}{35} \lcgs.

\subsection{SXP~327} \label{s:SXP327}
Pulsations from SXP~327 were first observed with \chandra in 2006
\citep{Laycock10}. \citet{Udalski08} reported an orbital period of
45.995 d using the MACHO and OGLE-III data of the optical
counterpart.  
The observed spin period (326.813 $\pm$ 0.031 s)
is consistent with the previous result by \citet{Laycock10}.
The X-ray flux of the source decreased by $\sim$ 10\%
between the two observations in 2006, which were a day apart (Figure~\ref{f:pdB}). 
The overall X-ray spectrum is fitted
by an absorbed power-law model with 
$\Gamma$ = 0.90 $\pm$ 0.03, and the observed X-ray luminosity
was about 10\sS{36} \lcgs.

The pulse profile shows
an eclipse-like dip in phases between 0.65 and 0.8.  The phase-resolved
quantile and the time versus phase diagrams
show marginal but complex spectral changes during pulsation cycles
(Figure~\ref{f:sxp327}). The spectral model fits of
the segmented phases summarized in Table~\ref{t:phrs_sxp327} indicate
that during the dip (phases 0.65--0.8) the spectrum becomes harder ($\Gamma$ $\sim$ 0.7)
with a less absorption (\nH $\sim$ 0) compared to the rest of the phases
($\Gamma$ $\ge$ 0.8, \nH $\ge$ 10\sS{21} cm\sS{-2}).

\subsection{SXP~504}
Pulsations of SXP~504 were independently discovered by
\citet{Edge04a} from the \chandra archival data and by \citet{Haberl04} 
from an \xmm observation. 
\citet{Edge04a} reported a period of 268.6 $\pm$ 0.1 d in the MACHO and
OGLE-II data of the optical counterpart.

SXP~504 was observed again with \chandra twice, five months apart, in 2013.
In both observations, the source was detected near a CCD edge and
the modulation period is close to the 1/2 harmonics of the known 1000 s
dithering period, but there was no sign of periodic modulations at 1000 s.
Both observations show a strong modulation at $\sim$ 501.3 sec, which
is consistent with the periods observed in the past \citep{Klus14}.
The overall X-ray luminosity increased by about 60\% between the two observations
in 2013 (from \nep{3.9}{34} to \nep{6.1}{34} \lcgs, Figure~\ref{f:pdB} and Table~\ref{t:spec}).

The X-ray spectrum of the first observation is well
fitted with an absorbed power-law, thermal bremsstrahlung, and APEC models,
whereas the second observation shows a good spectral fit only with an
absorbed power-law model. 
For an absorbed power-law model,
the X-ray spectra show similar photon indices ($\Gamma$ = 0.8 $\pm$
0.1, 0.60 $\pm$ 0.08) between the two observations 
whereas the absorption had dropped from \nH =
\nep{1.3}{22} to \nep{3}{21} cm\sS{-2}, which may explain the
increase in the observed X-ray luminosity of the second observation.
The pulse profiles also show minor changes
between the two observations, exhibiting more features (e.g., a sharp
rise at a phase around 0.6) and a larger modulation amplitude (from
65\% to 79\%) in the second observation (Figure~\ref{f:pdB} and
Table~\ref{t:src}). 

\begin{figure*} 
\begin{center}
\includegraphics*[width=0.90\textwidth,clip=true]{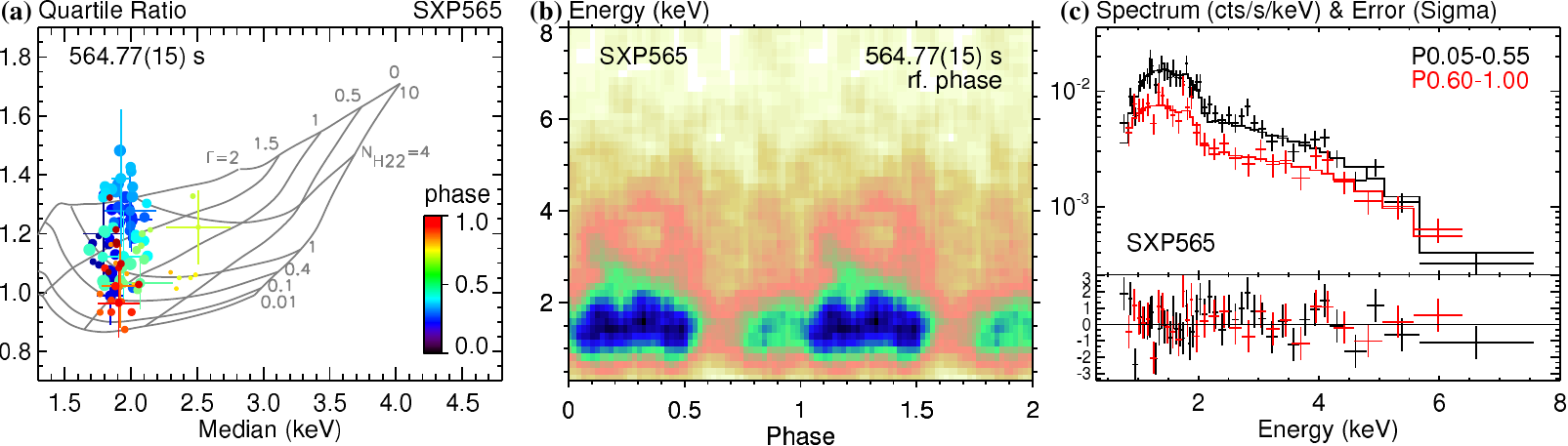}
\caption{Phase-resolved quantile diagram, energy versus phase diagram
and spectral model fits of SXP~565. See Section~\ref{s:SXP565}.}
\label{f:sxp565}
\end{center}
\end{figure*}

\subsection{SXP~565} \label{s:SXP565}
Pulsations from SXP~565 were first discovered 
from the 2001 \chandra observation by \citet{Macomb03}.
\citet{Schmidtke04} reported a period of 95.3 d from the
optical counterpart, while \citet{Edge05} did not detect this period
from the early OGLE data.  \citet{Galache08} reported a period of 151.8 d 
from the \rxte observations, and \citet{Rajoelimanana11} measured a period
of 152.4 d from the OGLE-III data, suggesting that the 152.4 d period is
the orbital period of SXP~565.

SXP~565 was observed again with \chandra in 2013 March and September.
The observed X-ray luminosity has increased to $\sim$\nep{2.4}{34} \lcgs 
in the 2013 observation from $\sim$\nep{1.9}{34} \lcgs in 2001.
The pulsation, however, was only detected in the 2001
observation although it was
observed far off-axis  in the ACIS S3
chip (11\arcmin\  from the aimpoint). The upper limit of the modulation amplitude is about 27\% in the 2013
observations, while the observed modulation amplitude in the 2001
observation is about 67\%. 

The X-ray spectrum became a bit harder in 2013 compared to 2001.
For an absorbed power-law model,
the absorption increased (\nH = \nep{1.0}{22} to \nep{2.8}{22} cm\sS{-2})
while the photon index remained constant ($\Gamma$ = 1.16 $\pm$ 0.07
to 1.07 $\pm$ 0.09, Table~\ref{t:spec}).
The pulse profile shows step-wise rises at phases $\sim$ 0.8 and 0.1 
from the minimum to maximum (Figure~\ref{f:pdC}). The
phase-resolved quantile and the energy versus phase diagrams in
Figure~\ref{f:sxp565} show a complex variation of the absorption, the
photon index, and the flux
during the pulsation. The spectral fits of the high and low state phases
in Figure~\ref{f:sxp565} and Table~\ref{t:phrs_sxp565} show that
the X-ray spectrum during the high flux phases (0.05--0.55) is softer
($\Gamma$ $\sim$ 1.3 versus 0.8) but more absorbed (\nH $\sim$
\nep{1.4}{22} versus \nep{5}{21} cm\sS{-2}) than the spectrum of the rest.

\subsection{SXP~701} \label{s:SXP701}
Pulsations from SXP~701 were first discovered from an \xmm observation
by \citet[][2004c]{Haberl04b}. \citet{Schmidtke05} and \citet{Rajoelimanana11}
found a 412 d period from the MACHO and OGLE-III data of
the optical counterpart, respectively.

SXP~701 was observed with \chandra in 2013 January and June.
The X-ray luminosity has dropped roughly by half between the two
observations from \nep{2.2}{35} \lcgs to 10\sS{35} \lcgs (Table~\ref{t:spec}). 
The X-ray spectra of the two observations are well fitted by an absorbed
power-law model. Both the photon index ($\Gamma$ $\sim$ 0.9 --
1.0) and the absorption (\nH $\sim$ \nep{2}{21} cm\sS{-2}) have remained more
or less constant between the two observations.

\begin{figure*} 
\begin{center}
\includegraphics*[width=0.90\textwidth,clip=true]
{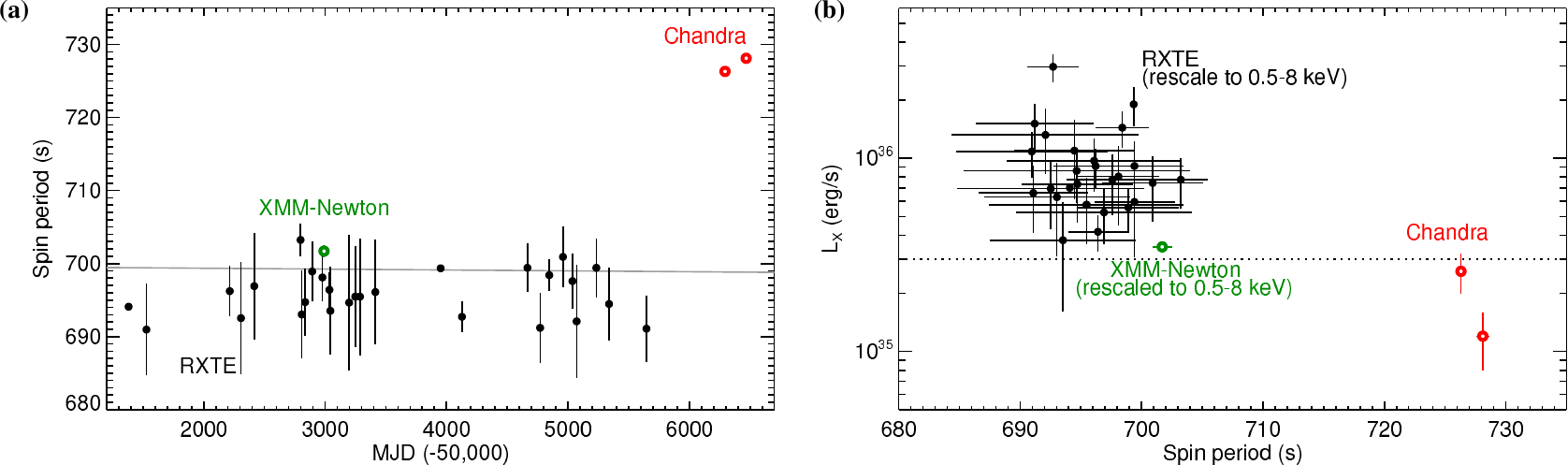}
\caption{{\bf (a)} Spin period evolution of SXP~701. The \rxte measurements
(black) and the
long-term trend (the grey line) are from \citep{Klus14}. The \xmm
(green) and \chandra
(red) measurements are from \citet{Haberl04c} and this analysis, respectively.
{\bf (b)} 
The 
intrinsic 0.5--8 keV X-ray luminosity ($L_X$) versus the spin period
of SXP~701. 
The X-ray luminosities from \rxte and \xmm are rescaled to the 0.5--8
keV band for an absorbed power-law model with $\Gamma$=1 and \nH =
\nep{2}{22} cm\sS{-2} (and also with a 60\% pulsed fraction for
\rxte: see Eq.~(1) in \citep{Klus14}).
}
\label{f:sxp701}
\end{center}
\end{figure*}

The source showed a strong pulsation at 726.31 $\pm$ 0.44 s and
728.11 $\pm$ 0.57 s in the 2013 January and June observations,
respectively.  As shown in Figure~\ref{f:sxp701}(a), these periods are
not consistent with the periods observed in the past \citep{Klus14}, which 
show essentially no long-term change 
($\dot{P}$ $\sim$ 0.0 $\pm$ 0.3 s yr\sS{-1}, $P_s$ $\sim$ 691--703 s). 
The pulsar appears to be experiencing a rapid spin-down trend ($\dot{P}$
$\sim$ $+$3.8 s yr\sS{-1}) between the two \chandra observations.
When comparing with the mean period ($P_s$ $\sim$ 699 s) or the last
data point ($P_s$ $\sim$ 691 s at $\sim$ MJD 55650) of the RXTE
measurements reported by \citet{Klus14}, the change appears even more
extreme with $\dot{P_s}$ $\sim$ $+$15 -- 20 s yr\sS{-1}.

In terms of the characteristic spin-down timescale ($\tau$\Ss{sd} =
$P/2\dot{P}$), the observed change in the spin period is remarkable:
$\tau$\Ss{sd} $\sim$ 20--100 yr. For comparison, 4U 2206$+$54, which is
an HMXB with a slowly spinning pulsar in a rapid spin-down trend
($P_s$ $\sim$ 5555 s, $\dot{P_s}$ $\sim$ 16 s yr\sS{-1}) 
shows $\tau$\Ss{sd} $\sim$ 180 yr
\citep{Wang12,Ikhsanov13}. 
According to the long-term spin trends of the SMC pulsars
discussed by \citet{Klus14}, only three pulsars (SXP~74.7, SXP~91.1, and SXP~342)
show a larger relative change in their spin periods over the 15 year
monitoring using \rxte, but their
spin periods are much shorter than SXP~701. SXP~756 and
SXP~1323, which may be similar to SXP~701 in terms of the binary parameters,
have been exhibiting large chaotic changes in their
spin periods.  However, what is unique about SXP~701
is its spin period has been somewhat stable until our survey.
Therefore, the recent change in the spin period of SXP~701 may provide a
new clue to the origin of the slowly spinning pulsars. 

Fig.~\ref{f:sxp701}(b) illustrates the
relationship between the X-ray luminosities and the spin periods of SXP~701. 
According to \citet{Shakura12}, who showed that a quasi-spherical
subsonic accretion flow can slow down the NS without the need for a high
surface magnetic field, a subsonic settling can be realized at relatively
low X-ray luminosities (i.e., low accretion rate), 
$L_X$ $<$ \nep{4}{36} \lcgs. While the precise threshold of the X-ray
luminosity for the transition may vary for
each system, Fig.~\ref{f:sxp701}(b) suggests that SXP~701 may have undergone a
similar transition at $\sim$ \nep{3}{35} \lcgs. 
This is reminiscent of torque reversals 
observed in the wind-fed pulsars like Vela X-1, GX 301-2, and GX 1+4,
but the precise distance information of the SMC pulsars allows an
accurate determination of the threshold luminosity, if any, (and thus
the corresponding accretion rate) for the
transition.  In fact, the suggested transition luminosity of
SXP~701 is similar to the turn over at $\sim$ \nep{4}{35} \lcgs in
the X-ray luminosity function of the SMC HMXBs (Antoniou et al.~2017
in preparation).  This implies that the spin-down trend of SXP~701 may
have been triggered by or related to the propeller effect \citep[the
centrifugal inhibition of accretion due to the pulsar's magnetic
field][]{Illarionov75}, perhaps suggesting yet another path to
slowly spinning pulsars.

\subsection{SXP~756} \label{s:SXP756}
SXP~756 was discovered as the slowest X-ray pulsar at the time by
\citet{Yokogawa00b}
from \asca observations. \citet{Laycock05} reported an X-ray period of
396 d, and \citet{Cowley03} and \citet{Schmidtke04} reported
recurring optical outbursts at $\sim$ 394 d intervals from the MACHO data of
the counterpart. \citet{Galache08} found an orbital period of
389.9 $\pm$ 7.0 d from the \rxte data.

SXP~756 was observed with \chandra in 2006. 
The pulsation period is found at 746.43 $\pm$ 0.22~s, which
agrees with the previous report by \citet{Laycock10} from the same
data set.  The 1/2 harmonic of the period also forms a significant peak
in the LS periodogram.  The X-ray spectrum is well fitted by an absorbed
power-law model with $\Gamma$ = 0.76 $\pm$ 0.05 and
\nH = \nep{1.1}{22} cm\sS{-2}.  The observed X-ray luminosity is
\nep{4.7}{34} \lcgs at 60 kpc.  The X-ray count rate in
Figure~\ref{f:pdC} shows a slight increase during the observations. 

\begin{figure*} 
\begin{center}
\includegraphics*[width=0.90\textwidth,clip=true]{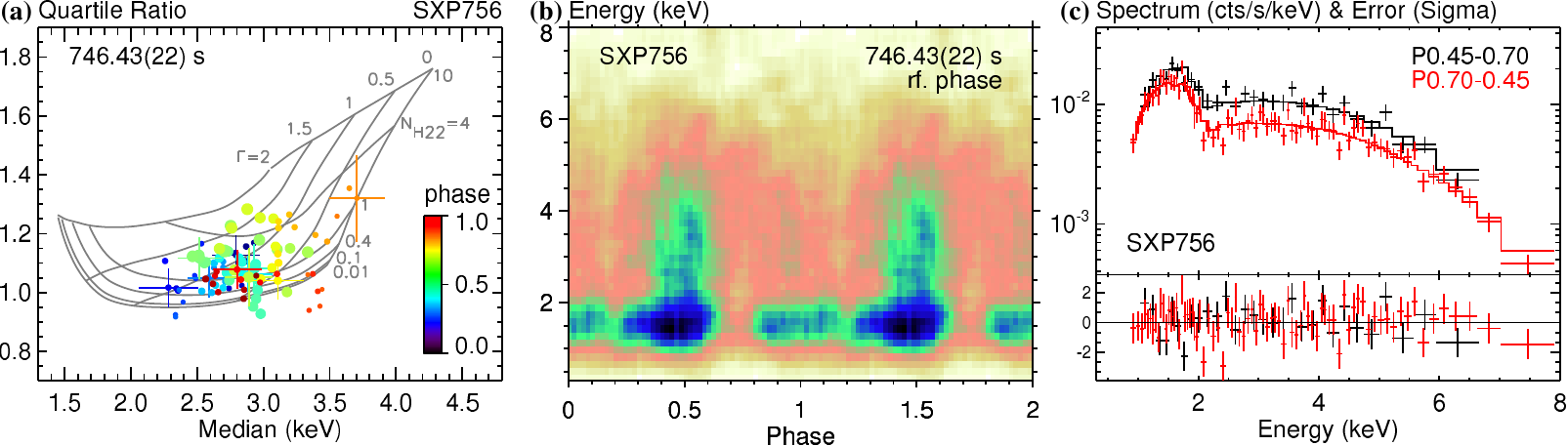}
\caption{
Phase-resolved quantile diagram, energy versus phase diagram and
spectral model fits of SXP~756.
See Section~\ref{s:SXP756}.}
\label{f:sxp756}
\end{center}
\end{figure*}

The phase-resolved quantile diagram in Figure~\ref{f:sxp756}(a)
reveals the spectral variations correlated with the pulsation cycle.
The energy versus phase diagram in Figure~\ref{f:sxp756}(b)
as well as the multiband folded light curves in Figure~\ref{f:pdC}
indicate separate minimum phases for soft and hard X-rays: at
phases $\sim$ 0.7 below 2 keV and at phases $\sim$ 0.2 above 2 keV.
The phase-resolved spectral fits in Figure~\ref{f:sxp756}(c) and
Table~\ref{t:phrs_sxp756} show that the high flux phase (0.45--0.70)
exhibits a heavier absorption by a factor of 2 while the photon index
remains constant ($\Gamma$ $\sim$ 0.8).

\subsection{SXP~893}
Pulsations from SXP~893 were first discovered 
from a 2001 \chandra observation by
\citet{Laycock10}.
Since then, SXP~893 was observed again with \chandra in 2013 March and
September.  The periodic X-ray
modulations are detected in the 2001 November and 2013 September
observations with $A$\Ss{mod} $\gtrsim$ 80\%.  The 2013 March
observation, when the X-ray count rate dropped by a factor of $>$ 2
compared to the other observations, does not show any significant periodic modulation
($A$\Ss{mod} $<$ 41\%).
The measured periods are within the normal
range of the spin periods observed with \rxte by \citet{Klus14}.
The pulse profile from the 2001 observation shows an eclipse-like dip
at phase $\sim$ 0.2, and the 2013 pulse profile shows a similar
dip in the flux at phase $\sim$ 0.5 (Figure~\ref{f:pdC}), but note the
different definition of phase 0.

The X-ray luminosity almost doubled between the 2001 November and 2013
September observations (from \nep{1.7}{35} to \nep{2.9}{35} \lcgs, Table~\ref{t:spec}).
Note that our estimate of the X-ray luminosity during the 2001 observation
is 80\% higher than the estimate by \citet{Laycock10}. The difference is 
due to the different assumptions of the spectral models and parameters.
The X-ray spectra are well fitted by an absorbed
power-law model.  While the photon index remained more or less constant
($\Gamma$ $\sim$ 0.6--0.7) between the two observations, the absorption
\nH dropped by more than half from \nep{1.6}{22} to \nep{7}{21}
cm\sS{-2} (albeit with large uncertainties).

\subsection{SXP~5.05} \label{s:SXP5.05}
SXP~5.05 was first discovered as a transient X-ray source with
\integral by \citet{Coe13a}.
Designated as IGR J00569--7226, it is
a relatively new addition among the SMC pulsars described in this paper.
Its pulsation was first discovered in a combination of \swift and \xmm
observations by \citet{Coe13b}. \citet{Coe15b} determined the orbital
parameters with a binary period of 17.13 $\pm$ 0.14 d through the
measurement of the changes in the spin periods from multiple \swift
observations.  They also argued that the NS is viewed
through the circumstellar disk periodically during orbital motions, which
leads to the observed flux variation aligned with the orbital period.

SXP~5.05 was observed with \chandra in 2001 January and 2013 January, June
and September.  Only in the last observation the source was detected
with a high count rate reaching 0.8 counts s\sS{-1}. 
The source was observed at 6.4\arcmin\  off the aimpoint with \chandra,
so there is no or little pile-up in the \chandra data.  
The upper limit (3$\sigma$) of the X-ray luminosity during the early
observations
with no detection ranges from 2.3 to \nep{4.7}{33} \lcgs at 60 kpc.

\begin{figure*} 
\begin{center}
\includegraphics*[width=0.80\textwidth,clip=true]{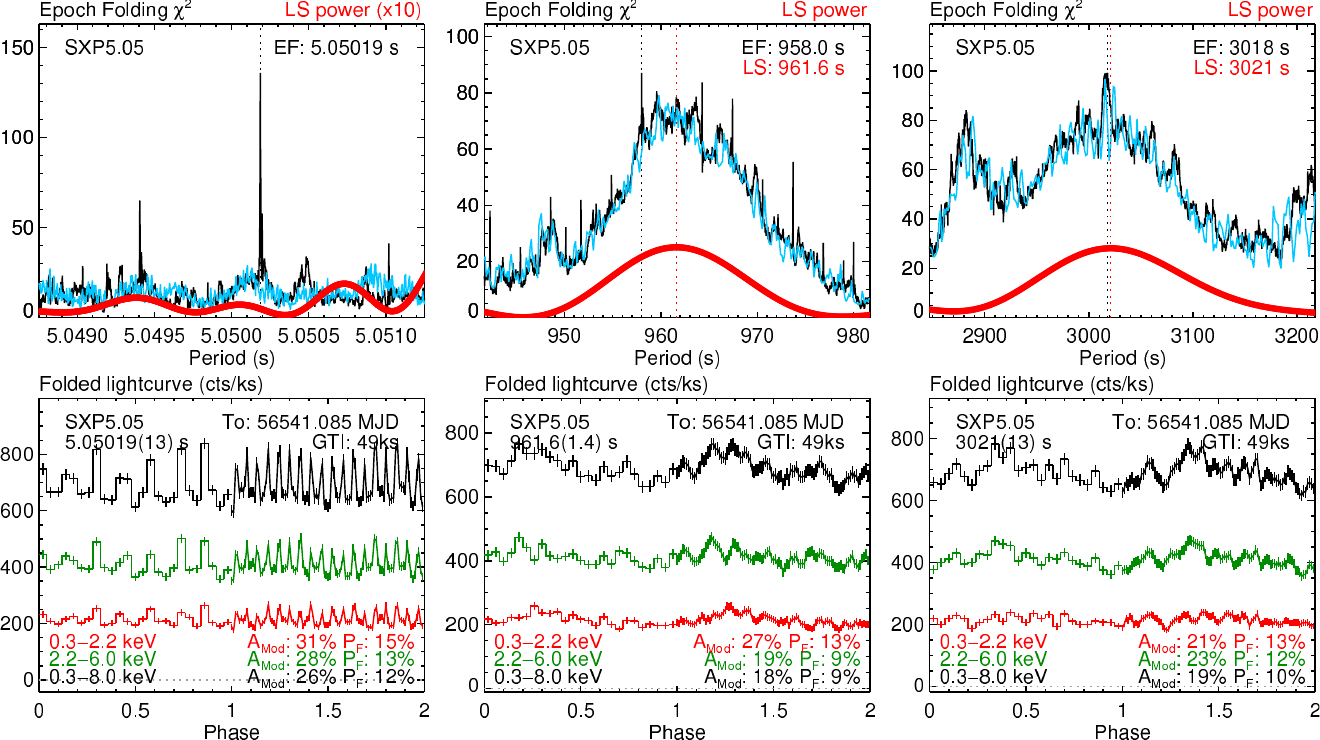}
\caption{
EF and LS periodograms ({\it top} panels)  and corresponding folded
light curves ({\it bottom} panels) in the
0.3--2.2, 2.2--6.0, and 0.3--8.0 keV bands of SXP~5.05 at around 5.05 s
({\it left}), 958 s ({\it middle}) and 3018 s ({\it right}) periods. The
folded light curve at 5.05019 s is choppy due to phase mixing and aliasing
from the similar CCD readout cycle (3.14 s). The light blue lines
indicate the EF periodogram using the time series randomized within
each readout cycle.  The EF folding periodogram is noisy near the 3018 s period, which is 
likely related to the long-term variation seen in the unfolded light curve
in Figure~\ref{f:pdD}.  See Section~\ref{s:SXP5.05}.}
\label{f:sxp5.05}
\end{center}
\end{figure*}

The pulsation period of SXP~5.05 is too short to be detected with the
\chandra ACIS chips using the LS periodogram, given the $\sim$ 3.14 s
readout cycle. The EF periodogram of the last observation, albeit noisy,
shows a sharp peak
at 5.05019 $\pm$ 0.00013 s (Figure~\ref{f:sxp5.05}), but it is likely
due to phase aliasing between the pulsation period and the CCD readout
cycle because (1) the pulse profile folded at 5.05019 s rapidly fluctuates
between the minimum and maximum during a pulsation cycle and (2) the
periodogram, which is generated from the randomized time tags of X-ray
events within each readout cycle, does not exhibit
sharp peaks (light blue curves in Figure~\ref{f:sxp5.05}).

The LS periodogram shows additional marginal periodicities at 958.0 $\pm$
1.4 s and 3018 $\pm$ 13 s (Figure~\ref{f:pdD}).  While the origin of
these periods is not clear, the light curve of the 2013 September
observation in Figure~\ref{f:pdD} exhibits several oscillations at
$\sim$ 8--10 ks period (4 BBs in Obs.~ID 15504). 

The \chandra light curve also shows an overall
decrease in the count rate from 800 to 600 counts per ks, indicating
that the observation may have caught a part of the long-term decline
from an outburst in 2013 \citep{Coe15b}.
On the other hand, the \xmm observations of SXP~5.05 in 2013 November
and December show $L_X$ $>$ 10\sS{37} \lcgs \citep{Coe15b}, which is
higher than $L_X$ $\sim$ \nep{9.1}{36} \lcgs measured with
\chandra in 2013 September, implying possible multiple outbursts from
the source.
The \chandra X-ray spectrum
is well fitted by an absorbed power-law model with 
$\Gamma$ = 0.17 $\pm$ 0.01.

\subsection{CXOU J003942.37--732427.4 or SXP~7.59?}
CXOU J003942.37--732427.4 is a new candidate pulsar
with a marginal detection (82.5\% confidence) of the pulsation at
7.58712 $\pm$ 0.00012 s from
the 2013 \chandra observation ($A$\Ss{mod} = 58\%). 
The source was observed again with \chandra in 2014 when its X-ray
luminosity increased to \nep{7}{34} \lcgs from \nep{3}{34} \lcgs 
in 2013. The X-ray count rates also doubled between the two
observations (Table~\ref{t:spec}), but the 2014 observation does not
show any periodicity ($A$\Ss{mod} $<$ 65\%).  
Given the small number
of total net counts (151) detected in the 2013 observation, further
observations in future will be needed to confirm the pulsation of
the source. 
Also note that there is no proper
counterpart for an HMXB at the \chandra position: the nearest
possible counterpart is an AGN \citep{Kozlowski13}.
The lack of early-type star in the error circle of the \chandra
position suggests that if it is a pulsar, it is not in an HXMB system.

According to the quantile analysis (Table~\ref{t:spec}), the X-ray
spectrum softened between the two observations 
due to an apparent reduction in the absorption from \nH $\sim$ \nep{7}{22} cm\sS{-2} to
\nep{2}{22} cm\sS{-2} with a constant photon index of
$\Gamma$ $\sim$ 1.2 for an absorbed power-law model.
The reduction in the absorption partially explains the increase in the observed
X-ray luminosities between the two observations.

\subsection{SXP~8.02} \label{s:sxp8.02}

SXP~8.02 is the only known magnetar in the SMC. \citet{Lamb02b} first
suggested that SXP~8.02 is an anomalous X-ray pulsar (AXP) based on the
detection of X-ray pulsations from the 2001 \chandra observation. 
The fast Fourier transform (FFT) analysis by \citet{Lamb02b} showed
roughly the equal power from the 5.4 and 8.02 s periods.
Our LS periodogram shows that the main period at 8.018944 $\pm$
0.000058~s has a much stronger power ($X$ $\sim$ 18) than the 5.43988~s
period ($X$ $\sim$ 12).
Unlike the double peak profile seen in the \xmm
observations \citep{Tiengo08}, the \chandra pulse profile 
shows a single broad peak over
each pulsation cycle (Figure~\ref{f:pdD}), which could be in part due
to phase mixing from the 3.24~s CCD readout cycle.

The X-ray spectrum of SXP~8.02 is the softest among the pulsars in this survey and
softer than the majority of the SMC field X-ray sources (see
Section~\ref{s:discussion}). It cannot
be fitted by any of the four basic spectral models considered 
here: e.g., \citet{Lamb02b}
got $\chi^2_r$ $\sim$ 1.4 for an absorbed blackbody model. 
Instead, it is better fitted ($\chi_r^2$ $\sim$ 1.15) by an absorbed
two-temperature blackbody model with $kT$s of 0.39 $\pm$ 0.06 and 3.7
$\pm$ 0.6 keV. 
The X-ray spectrum is consistent with those reported by \citet{Tiengo08}
from the \xmm observations. \citet{Tiengo08}
constrained the NS radius at 12.5 km using a two-temperature blackbody model.
The 0.5--8 keV X-ray luminosity is estimated to be \nep{2.0}{35} \lcgs,
which is somewhat higher than \nep{1.3}{35} \lcgs by \citet{Lamb02b}.
The difference is likely due to the different assumptions of the
spectral model.

\subsection{SXP~74.7} \label{s:SXP74.7}
X-ray pulsations from SXP~74.7 were first discovered with
\rxte by \citet{Yokogawa98}. \citet{Schmidtke05} reported an orbital
period of 33.4 $\pm$ 0.4 d in the MACHO light curve, and more recently
\citet{Rajoelimanana11} reported a period of 33.38 $\pm$ 0.01 d in
the detrended light curve generated from the MACHO and OGLE observations.

SXP~74.7 was observed with \chandra in 2013 and 2014. 
The X-ray luminosity decreased from \nep{1.7}{35} \lcgs
to \nep{3.0}{34} \lcgs between the two observations. The X-ray
spectrum is well fitted by an
absorbed power-law model with $\Gamma$ $\sim$ 1 and no
significant spectral change is observed between the two observations.

\begin{figure*} 
\begin{center}
\includegraphics*[width=0.90\textwidth,clip=true]{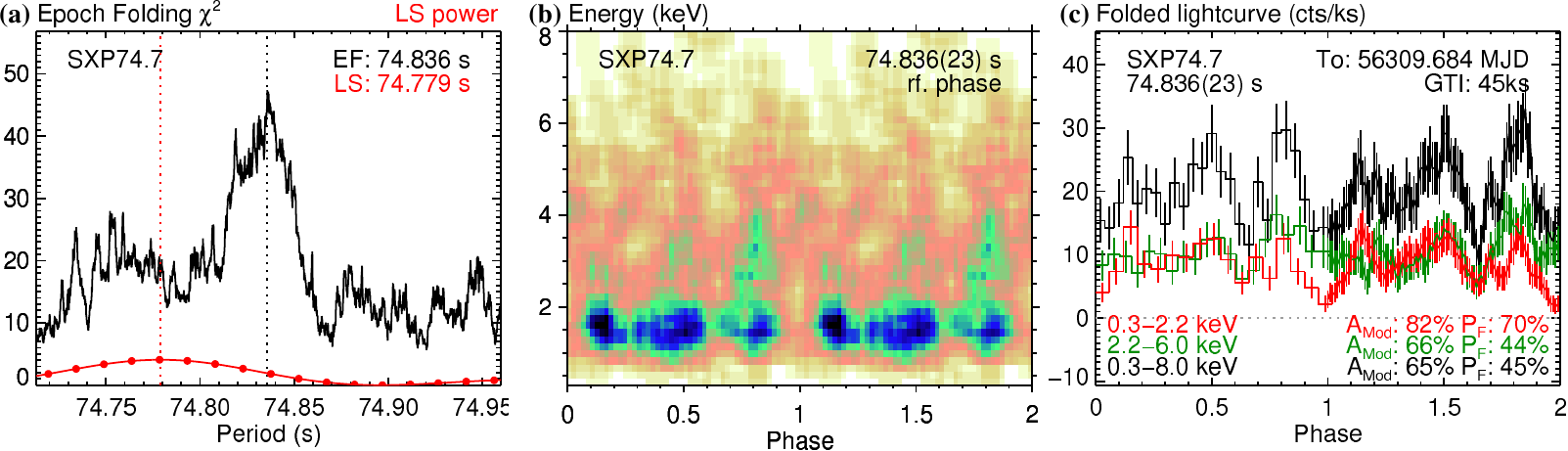}
\caption{
LS and EF periodograms,  energy versus phase diagram,
folded light curves ({\it right-middle}) 
of SXP~74.7 at the 74.836 s period.  See Section~\ref{s:SXP74.7}.
}
\label{f:sxp74.7}
\end{center}
\end{figure*}

The LS periodogram in Figure~\ref{f:pdD} shows a marginal periodicity
at 24.9452 $\pm$ 0.0012 s, but not at the reported
pulsation period of $\sim$ 74.7 s. 
On the other hand, the EF periodograms in Figure~\ref{f:sxp74.7} show
similarly significant peaks at both 24.9452 and 74.836~s.
The folded light curves show a single broad peak for the 24.9452~s
period and three distinct peaks for 74.836~s, which indicates that
the latter is the true spin period (similarly to SXP~51.0; see Section
\ref{s:SXP51.0}).  

The phase-resolved spectral analysis in Figure~\ref{f:sxp74.7} shows a
marginal spectral variation during pulsation: the X-ray spectrum gets
intrinsically harder with $\Gamma$ $\sim$ 2, 1.5, and 1 at the first,
second and third peaks, respectively.

\subsection{CXOU J005446.38--722523.0 or SXP~4693?} \label{s:sxp4693}

CXOU J005446.3--722523 was observed with \chandra in 2006
and 2013.  The source was detected in 2006, but not in 2013.
The observed X-ray luminosity was \nep{1.2}{34} \lcgs at 60 kpc and the upper
limit (3$\sigma$) during the 2013 observation was \nep{2}{33} \lcgs.
The X-ray spectrum is consistent with an absorbed power-law model
of $\Gamma$ $\sim$ 2 according to a quantile analysis.

The 2006 observation of the source shows a very long pulsation
period at 4693 $\pm$ 20 s \citep{Laycock10}.  Our analysis also shows
a marginal detection of the long pulsation at 4685 $\pm$ 23 s based on
the refined search using the LS periodogram.  Given the increase in the
period search trials in this analysis compared to \citet{Laycock10}, now
the LS power of the observed period corresponds to only about 50\%
confidence under the assumption of a blind survey. Note that the
total net counts of the source is only 151.

Intriguing is the fact that
CXOU J005446.3--722523 is about 2\arcsec\ off the reported optical
counterpart of SXP~6.88 \citep{Coe15}.
The EF periodogram
of CXOU J005446.3--722523 around 6.88 s does not show any clear sign
of periodicity above the noise distribution.  Since SXP~6.88 is an
\integral source with a relatively large positional uncertainty of the
X-ray source \citep{McBride07}, it is not clear that the observed long
pulsation from CXOU J005446.3--722523 is actually from
SXP~6.88.  There is also indication for the long period in \xmm data
and no indication for 6.88 s \citep{Haberl16}.

While additional observations of CXOU J005446.3--722523
are needed to validate the observed long spin period,
the origin of the slowly spinning pulsars have been challenging to the
conventional formation and evolutionary theory of pulsars, where long
spin periods ($P_s$ $\gtrsim$ 500 s) are associated with
extremely high magnetic field ($\gtrsim$ 10\sS{13} G) 
on the surface of the pulsar \citep[e.g.,][]{Ghosh79,Kluzniak07}. For instance, \citet{Finger10}
interpreted 4U 2206$+$54 ($P_s$ $\sim$ 5560 s) as an accreting
magnetar. Efforts to explain the long period pulsars by alternative
models are accumulating: \citet{Ikhsanov07} incorporated
a subsonic propeller state into 
the evolutionary tracks of NSs, which can overcome the spin period 
barrier of $\sim$ 500 s; \citet{Wang12} proposed a retrograde wind
accretion to explain long spin periods of supergiant fast X-ray transients
\citep[see also][]{Christodoulou17};
\citet{Ikhsanov13} suggested a magnetized accretion stream to explain
the long spin period of 4U 2206$+$54 where the magnetic field at the
surface of the NS is expected to be $\sim$\nep{4}{12}~G; 
\citet{Shakura12,Shakura13}
developed a model for quasi-spherical subsonic accretion, which can
explain long spin periods of the NSs in symbiotic XRBs.

%%text_stop

\begin{figure*} \begin{center}
\includegraphics*[width=0.90\textwidth,clip=true]{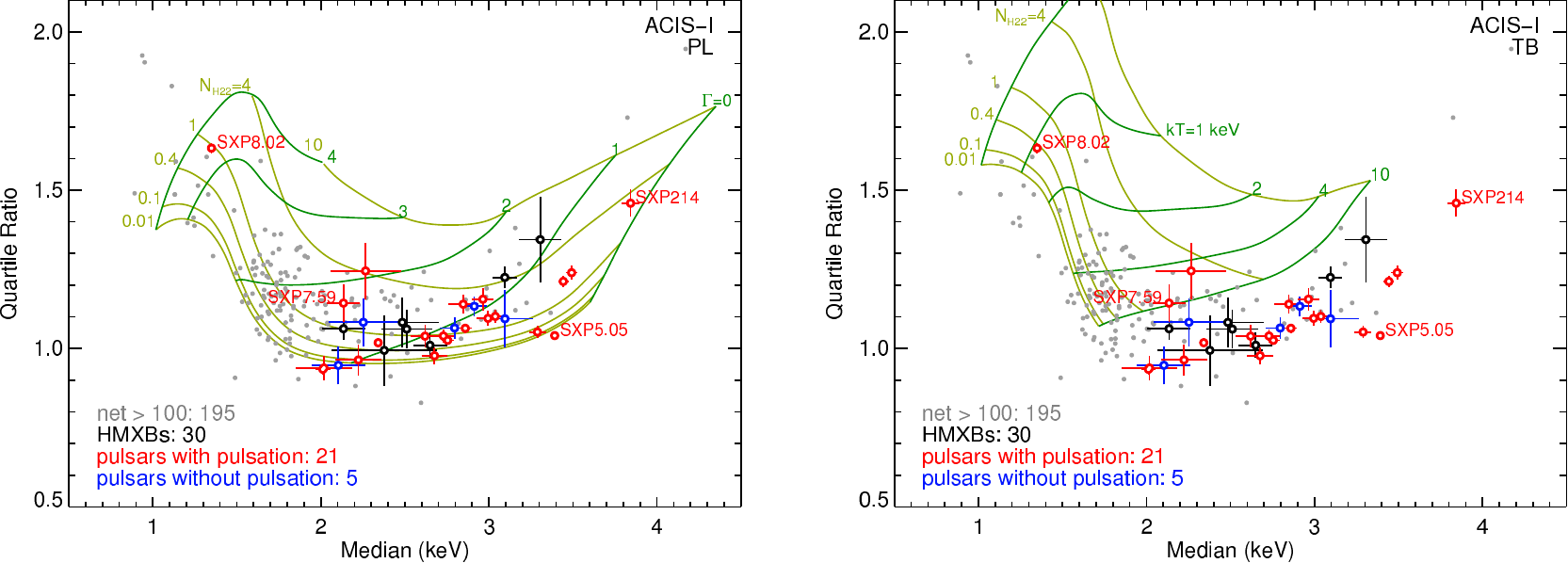}
\caption{Quantile diagram of the SMC pulsars (red and blue) and HMXBs
(black) in comparison with the SMC field sources (grey) with $\geq$ 100
net counts in 0.3--8 keV.  Only the sources detected in the ACIS-I
chips are shown.  Absorbed power-law ({\it left}) and thermal
bremsstrahlung ({\it right}) models are shown in grids for comparison. 
The grids cover the SMC local
absorption with \nH = 0.01, 0.1, 0.4, 1, 4, and 10 $\times$
10\sS{22} cm\sS{-2}.
In the power-law models, the grids cover
$\Gamma$ = 0, 1, 2, 3, and 4. In the thermal bremsstrahlung models,
the grids show the plasma temperature $kT$ = 0.4, 1, 2, 4, and 10 keV.
}
\label{f:qd}
\end{center}
\end{figure*}

\begin{figure*} \begin{center}
\includegraphics*[width=0.90\textwidth,clip=true]{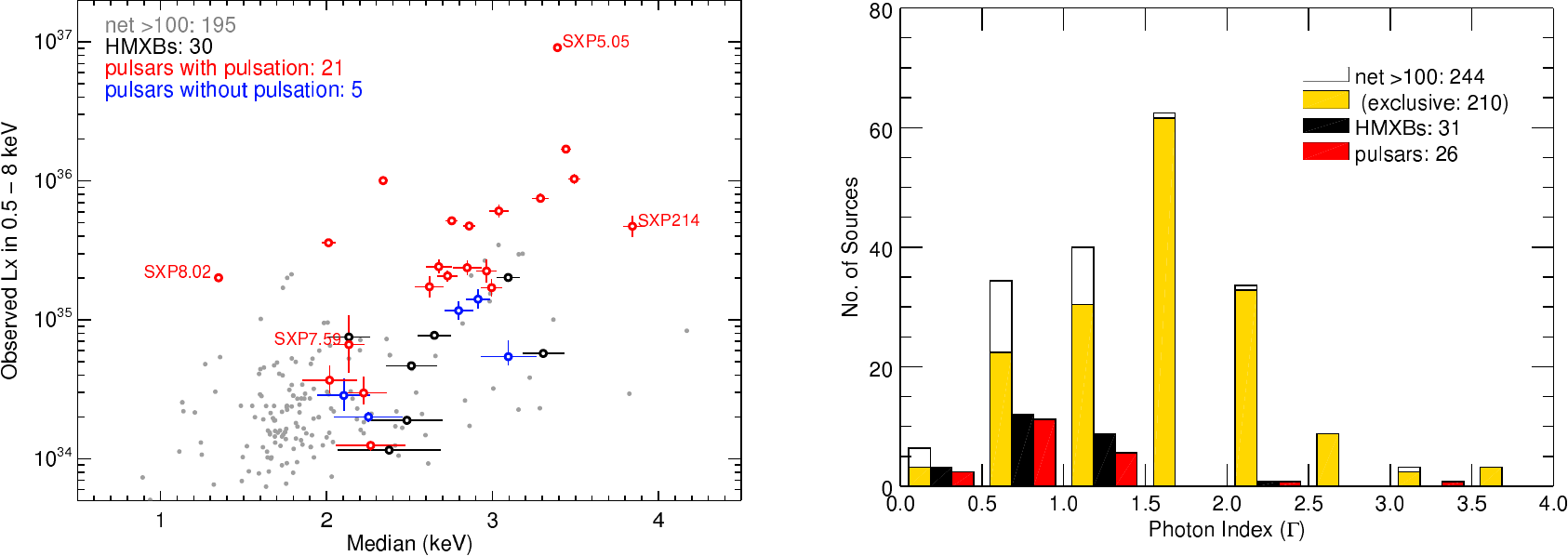}
\caption{Observed X-ray luminosity in 0.5--8 keV at 60 kpc versus
the median energy of the SMC field sources, HMXBs, and pulsars ({\it
left}, only from the ACIS-I chips) shown in Figure~\ref{f:qd}. 
Distribution of the photon indices for the absorbed power-law models
({\it right}, from both the ACIS-I and S chips). The X-ray spectra of
the SMC HMXBs and pulsars are in general harder than those of the SMC
field sources. 
}
\label{f:hist}
\end{center}
\end{figure*}

%%text_start
\section{Discussion} \label{s:discussion}

We have analyzed the deep \chandra survey of the SMC in search for
X-ray pulsars. 
The LS and EF analysis showed that 21 sources including
a new candidate pulsar (CXOU J003942.37--732427.4) exhibited X-ray
pulsations. We have also detected another 12 known pulsars with no X-ray
pulsations, six of which, though, were detected only with less than 100
net counts, and thus were
too faint for pulsation search. \citet{Laycock10} claimed 
pulsation detections from SXP~8.88 and SXP~15.3, but the claimed peaks of their
LS periodograms are well below the detection threshold set
in this analysis, so we do not include them in the list of the pulsation
detections.  It is possible that another four pulsars (SXP~4.78, SXP~6.88,
SXP~16.6, and SXP~95.2) were in the FoV of the \chandra observations.
SXP~6.88 could be the same source as SXP~4693 (Section~\ref{s:sxp4693}).  
The X-ray positions of the other three sources are too uncertain to conclusively
associate or rule out any association with \chandra sources.

Here we review the collective X-ray timing and spectral
properties of SMC pulsars, and compare them with those of HMXBs without
pulsations and the field sources in the SMC. See also \citep{Yang17}.

\begin{figure} 
\begin{center}
\includegraphics*[width=0.433\textwidth,clip=true]{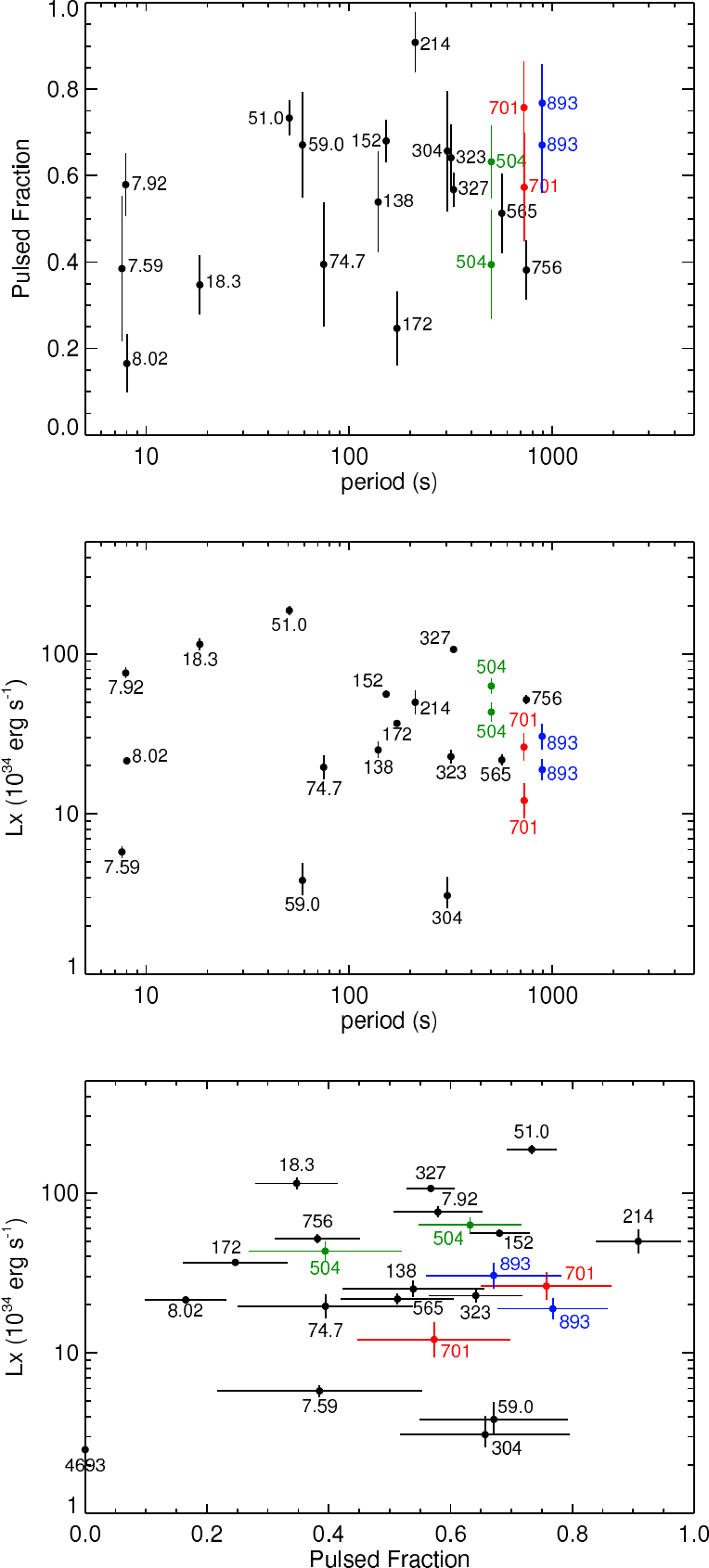}
\caption{Relations among the spin periods, the pulsed fractions, and the
observed 0.5--8 keV X-ray luminosities of the SMC pulsars (excluding SXP~5.05).
For SXP~504 (green), SXP~701 (red), and SXP~893 (blue), the results of
two measurements per each are shown. 
}
\label{f:pfpdlx}
\end{center}
\end{figure}

\subsection{X-ray properties of SMC pulsars and field sources}

Figure~\ref{f:qd} compares the energy quantiles of the SMC field sources
with the pulsars and HXMBs for absorbed power-law and thermal
bremsstrahlung models. The sources with at least 100 net
counts are plotted to limit the statistical uncertainty.
Only the sources detected in the ACIS-I chips are shown for 
a proper comparison with the spectral model grids since the grid pattern
depends on the detector response.
The spectral models assume two absorption
components: the fixed Galactic foreground and variable SMC local components
(Section~\ref{s:pulsars}), where the latter is parameterized
in the grid pattern along with the photon indices for the power-law
models 
and the plasma temperatures for the thermal bremsstrahlung models. 
All the pulsars but the new candidate (CXOU J003942.37--732427.4) and the magnetar
SXP~8.02 are identified as HMXBs as well.
Excluding the magnetar, the pulsars and the HMXBs exhibit intrinsically 
harder X-ray spectra (median energy $\gtrsim$ 2 keV) than the SMC field sources.
For the thermal bremsstrahlung models, the plasma temperatures of
$>$ 90\% of pulsars and HXMBs are above 10 keV.

Figure~\ref{f:hist} shows the observed X-ray luminosities
versus the median energy of the X-ray spectra.
All the \chandra sources with $L_X$ $\gtrsim$ \nep{4}{35} \lcgs exhibit 
X-ray pulsations, and more luminous pulsars show harder spectra. 
The majority of the field sources
are at luminosities below 10\sS{35} \lcgs, many of which 
have the median energy below 2 keV unlike the HMXBs and pulsars.
Figure~\ref{f:hist} also shows the distribution of the photon indices:
the field source distribution peaks at $\Gamma$ = 1.5--2.0, whereas
the distributions of the SMC HMXBs and pulsars peak at $\Gamma$ =
0.5--1.0. The observed photon indices of the SMC pulsars are in excellent
agreement with typical spectra of pulsars. The hard X-ray spectra of
HMXBs without known pulsation reinforce the notion that they are likely
associated with accreting pulsars.

\subsection{Spin periods, pulsed fraction and X-ray luminosities}

Figure~\ref{f:pfpdlx} shows the relations among the spin periods, the pulsed
fractions, and the observed 0.5--8 keV X-ray luminosities 
(during the pulsations) of the SXPs in our survey.
There appears to be no clear correlation among these observables.
The pulsed fractions and spin periods show a marginal correlation, but
the relatively low pulsed fraction at spin periods shorter than 10 s could
be in part due to the relatively long CCD readout cycle.
SXP~5.05 is excluded in Figure~\ref{f:pfpdlx} since its pulsed
fraction at the primary period is not reliably measured for its
spin period due to the CCD readout time being similar to the spin period. 

\subsection{Unusual spin periods}

\citet{Klus14} reported the long-term trend of the spin periods of 42
SMC pulsars using  the 15 yr \rxte data from 1997 to 2012. 
They found that 26 and 15 pulsars show spin-up and spin-down trends,
respectively, and one pulsar (SXP~701) shows no change on average.
Overall the average rate change of the spin periods is small with 
$\dot{P}/P$ ranging from \nep{-5}{-3} yr\sS{-1} to \nep{3}{-3} yr\sS{-1}, suggesting
that many of them might have achieved spin equilibrium.  
On the other hand, the NSs in high eccentric orbits of Be-XRB systems
can go through a periodic accretion phase, which can lead to a sudden
change in the spin period. 

In our analysis, three pulsars (SXP~51.0, SXP~214 and SXP~701)
show a large deviation in their spin periods compared to their previously
reported values or long-term trends. The eccentricities of these
pulsars are not known.
The recent spin-up trend of SXP~214 could be as high as $-\dot{P}/P$ $\sim$
\nep{8}{-3} yr\sS{-1}. See \citet{Hong16} for the details.
In the case of SXP~51.0, its spin-up trend appears to have doubled recently
($-\dot{P}/P$ $\sim$ \nep{4}{-3} yr\sS{-1}, Section~\ref{s:SXP51.0}). 
SXP~701 shows the most dramatic change in the spin period
with a rapid spin-down trend of $\dot{P}/P$ $\sim$ \nep{5.4}{-3} yr\sS{-1}
to $\sim$ \nep{2}{-2} yr\sS{-1}, breaking from a spin-equilibrium
state of likely more than 15 yrs (Section~\ref{s:SXP701}). If the
recent change is more permanent than chaotic, 
SXP~701 may provide a unique insight to the physics behind the
formation of slowly spinning pulsars.
Given the unusual rapid change in recent years, further monitoring of
these sources is needed to understand the cause of the spin change trend.

\subsection{Eclipse-like dips in pulse profiles}

The relatively low background in the \chandra observations due to the
superb angular resolution of the \chandra X-ray optics enables measurements of
precise pulsation profiles, which can be used for modeling the geometry of
the NS and the pulsated X-ray emissions. Six pulsars (SXP~51.0,
SXP~152, SXP~214, SXP~323, SXP~327, SXP~504) along with
SXP~565 and SXP~894 in some observations exhibit almost an
eclipse-like sharp dip in their folded light curves (Figures~\ref{f:pdA}--\ref{f:pdD}). Such distinct features in
the pulse profiles can be used to refine the period measurement
(Section~\ref{s:refine}).

\subsection{Spectral changes during pulsation cycles}

Phase-resolved spectral analysis using quantile and energy versus phase
diagrams reveals that 
many SXPs exhibit diverse spectral variations over the pulsation
cycles. Two outstanding pulsars in this respect are SXP~172 in Figure~\ref{f:sxp172}
and SXP~214 \citep{Hong16}. In the case of SXP~172, the quantile diagram
shows that the overall photon indices vary significantly for an absorbed power-law model,
and the energy versus phase diagram reveals that the dominant pulsation
component comes from hard X-rays above 2 keV.  In addition, SXP~18.3
and SXP~756 exhibit changes in the absorption during the
pulsation cycles, whereas SXP~152 and SXP~304 show an intrinsic spectral
variation, i.e., changes in the photon indices for absorbed power-law
models.  SXP~51.0, SXP~327 and SXP~565 show changes in both the absorption and
photon index during the pulsation cycles.

The diverse pulse profile and spectral changes during pulsation phases
observed in this analysis provide a rich data set for modeling the NS
geometry and X-ray emission \citep[e.g.,][]{Sasaki12}.

\section{Acknowledgment}

VA acknowledges financial support from NASA/Chandra grants GO3-14051X,
AR4-15003X, NNX15AR30G and NASA/ADAP grant NNX10AH47G.  AZ acknowledges
funding from the European Research Council under the European Union's
Seventh Framework Programme (FP/2007-2013)/ERC Grant Agreement n.~617001
and the MSCA-RISE-GA-691164 "ASTROSTAT" progrmme.  PP acknowledges
financial support from NASA contract NAS8-03060.  MS acknowledges support
by the the Deutsche Forschungsgemeinschaft through the research grant (SA
2131/4-1) and the Heisenberg professorship (SA 2131/5-1).  SL acknowledges
financial support from NASA/ADAP grant NNX14-AF77G.

%%text_stop

%%text_stop
\clearpage

\begin{sidewaystable}
\vspace{7cm}
\scriptsize
\caption{SMC pulsars}
\input{tab3.tex}
\label{t:src}
Notes --
(1) Pulsar Name.
(2) \chandra source Name.
(3) Observed pulsation period. ${}^\dagger$ indicates the refined periods by the EF method. 
(4) Orbital periods of the system \citep{Coe15}.
(5) The peak value of the LS periodogram.
(6) False detection probability corresponding to the LS power.
(7) Modulation Amplitude.
(8) Pulsed fraction.
(9) Net counts in 0.3--8 keV.
(10) Median energy value of the source photons in 0.3--8 keV.
(11) Offset between the \chandra source and the pulsar position
reported in \citet{Coe15} for SXP~8.02 and SXP~9.13 or
\citet{Haberl16} for the rest.
(12) Number of individual observations, source detections and pulsation detections.
(13) Same as (12) for the stacked observations.
(14) Start date of the observation.
(15) Good time intervals (GTIs).
(16) Flags. "a": absorption variation during pulsation cycle, "i": intrinsic spectral variation during
pulsation cycle, "e": eclipse-like feature in pulse profile, "v": flux variation.
(17) Observation and Deep Field IDs. The numbers in parentheses
represent 1$\sigma$ equivalent errors.
\end{sidewaystable}
\clearpage

\begin{sidewaystable}
\vspace{7cm}
\scriptsize
\caption{Spectral properties}
\input{tab4.tex}
\label{t:spec}
Notes --
(1) Pulsar Name.
(2) Net counts in 0.3--8 keV.
(3) Mode of the posterior probability for Bayesian Enhanced Hardness Ratio (BHER).
(4) Median energy value of the source photons in 0.3--8 keV.
(5) Quartile ($E_{25\%}$ and $E_{75\%}$) ratio.
(6) Quantile diagram based photon index ($\Gamma$) and
(7) Extinction (\nH in 10\sS{22} cm\sS{-2}) estimate under an absorbed power-law model (${}^\dagger$ for an absorbed thermal bremsstrahlung model).
(8) Spectral model fit based photon index ($\Gamma$) and
(9) Extinction (\nH in 10\sS{22} cm\sS{-2}) estimate under an absorbed power-law model (${}^\ddag$ for an absorbed two-temperature blackbody model).
(10) Reduced $\chi^2$ and Degrees of Freedom (DoF) of the fit.
(11) Flux estimate in 0.5--2 keV and (12) in 2--8 keV.
(13) Observed and (14) intrinsic luminosity estimates at 60 kpc.
(14) Observation and Deep Field IDs.
\end{sidewaystable}
\clearpage

\begin{table}
\tiny
\begin{minipage}{0.47\textwidth}
\caption{Phase resolved spectral analysis: SXP~18.3}
\input{tab5.tex}
\label{t:phrs_sxp18.3}
Notes.-- (4) in 10\sS{22} cm\sS{-2}. 
(6) the observed and (7) intrinsic 0.5--8 keV X-ray luminosities in 10\sS{35} \lcgs at 60 kpc. 
\end{minipage}
\end{table}

\begin{table}
\tiny
\begin{minipage}{0.47\textwidth}
\caption{Phase resolved spectral analysis: SXP~51.0}
\input{tab6.tex}
\label{t:phrs_sxp51.0}
Notes.-- (4) in 10\sS{22} cm\sS{-2}. 
(6) the observed and (7) intrinsic 0.5--8 keV X-ray luminosities in
10\sS{35} \lcgs at 60 kpc. 
\end{minipage}
\end{table}

\begin{table}
\tiny
\begin{minipage}{0.47\textwidth}
\caption{Phase resolved spectral analysis: SXP~152}
\input{tab7.tex}
\label{t:phrs_sxp152}
Notes.-- (4) in 10\sS{22} cm\sS{-2}. 
(6) the observed and (7) intrinsic 0.5--8 keV X-ray luminosities in 10\sS{35} \lcgs at 60 kpc. 
\end{minipage}
\end{table}

\begin{table}
\tiny
\begin{minipage}{0.47\textwidth}
\caption{Phase resolved spectral analysis: SXP~172}
\input{tab8.tex}
\label{t:phrs_sxp172}
Notes.-- (4) in 10\sS{22} cm\sS{-2}. 
(6) the observed and (7) intrinsic 0.5--8 keV X-ray luminosities in 10\sS{35} \lcgs at 60 kpc. 
\end{minipage}
\end{table}

\begin{table}
\tiny
\begin{minipage}{0.47\textwidth}
\caption{Phase resolved spectral analysis: SXP~327}
\input{tab9.tex}
\label{t:phrs_sxp327}
Notes.-- (4) in 10\sS{22} cm\sS{-2}. 
(6) the observed and (7) intrinsic 0.5--8 keV X-ray luminosities in 10\sS{35} \lcgs at 60 kpc. 
\end{minipage}
\end{table}

\begin{table}
\tiny
\begin{minipage}{0.47\textwidth}
\caption{Phase resolved spectral analysis: SXP~565}
\input{tab10.tex}
\label{t:phrs_sxp565}
Notes.-- (4) in 10\sS{22} cm\sS{-2}. 
(6) the observed and (7) intrinsic 0.5--8 keV X-ray luminosities in 10\sS{35} \lcgs at 60 kpc. 
\end{minipage}
\end{table}

\begin{table}
\tiny
\begin{minipage}{0.47\textwidth}
\caption{Phase resolved spectral analysis: SXP~756}
\input{tab11.tex}
\label{t:phrs_sxp756}
Notes.-- (4) in 10\sS{22} cm\sS{-2}. 
(6) the observed and (7) intrinsic 0.5--8 keV X-ray luminosities in 10\sS{35} \lcgs at 60 kpc. 
\end{minipage}
\end{table}

\end{document}

%% file: tab1.tex
\begin{tabular*}{\textwidth}{rcccc@{\extracolsep{\fill}}rccccl}
\hline\hline
                          & (1) & (2)        & \multicolumn{2}{c}{                } &                                &          & (3)  & (4)          & (5)           & (6)      \\
Obs.                      & DF  & Stacked    & \multicolumn{2}{c}{Pointing (J2000)} & \multicolumn{1}{c}{Start Time} & Exposure & GTI  & \multicolumn{3}{c}{Source Count} \\
\cline{4-5}\cline{9-11}ID & ID  & for Period & R.A.             & Dec                        &                                &          &      & Total        & Net $\ge$ 100 & Periodic \\
                          &     & Search     & (h m s)          & (\Deg\  \arcmin\  \arcsec) & \multicolumn{1}{c}{(UT)}       & (ks)     & (ks) & 0.5--7 keV   & 0.3--8 keV    & Sources  \\
\hline
1881                      &     &            & 00 59 05.0       & -72 10 42.1                & 2001-05-15 01:54               & 100.0    & 98.7 & 136          & 28            & 4        \\[0.15cm]
7155                      & 01A & 1          & 00 53 34.5       & -72 26 43.2                & 2006-04-25 05:15               & 50.0     & 49.3 & 109          & 14            & 2 (2)    \\
7327                      & 01A & 1          & 00 53 34.5       & -72 26 43.2                &     -04-26 14:57               & 50.0     & 47.4 & 100          & 8             & 3 (2)    \\[0.15cm]
8479                      & 02A & 2          & 00 50 41.4       & -73 16 10.3                &     -11-21 12:03               & 45.0     & 42.1 & 68           & 10            & 4 (3)    \\
7156                      & 02A & 2          & 00 50 41.4       & -73 16 10.3                &     -11-22 18:48               & 39.0     & 38.7 & 77           & 9             & 3 (3)    \\
8481                      & 02A & 2          & 00 50 41.4       & -73 16 10.3                &     -11-23 15:45               & 16.0     & 16.0 & 43           & 6             & 1 (1)    \\[0.15cm]
14666                     & 03  &            & 01 13 56.9       & -73 20 34.0                & 2012-12-28 14:49               & 50.0     & 49.4 & 79           & 12            &          \\[0.15cm]
14668                     & 05  &            & 00 56 08.6       & -72 35 02.5                & 2013-01-02 12:22               & 50.0     & 49.4 & 92           & 10            & 1 (1)    \\
14670                     & 07  &            & 00 51 52.0       & -73 00 24.6                &     -01-06 17:52               & 50.0     & 49.4 & 75           & 5             & 1        \\
14671                     & 08  &            & 00 56 35.9       & -72 20 06.3                &     -01-07 08:18               & 50.0     & 48.1 & 94           & 19            &          \\
14667                     & 04  &            & 01 13 39.9       & -73 08 37.2                &     -01-15 09:43               & 50.0     & 46.5 & 56           & 5             &          \\
14672                     & 09  &            & 00 49 43.2       & -72 49 16.3                &     -01-17 16:09               & 50.0     & 44.8 & 77           & 10            & 1        \\
14673                     & 10  &            & 00 41 00.0       & -73 20 00.0                &     -01-18 08:02               & 50.0     & 44.8 & 86           & 5             & 1        \\[0.15cm]
14674                     & 11  &            & 00 47 09.6       & -73 07 33.0                &     -03-04 22:33               & 50.0     & 45.9 & 71           & 10            & 2 (1)    \\
14665                     & 02  &            & 01 13 33.2       & -72 32 41.8                &     -03-09 05:11               & 50.0     & 47.8 & 89           & 7             &          \\
14669                     & 06  &            & 00 52 46.6       & -72 42 11.8                &     -03-11 07:15               & 50.0     & 47.4 & 65           & 4             &          \\
14664                     & 01  &            & 01 11 10.1       & -72 44 03.5                &     -03-25 02:20               & 50.0     & 49.4 & 75           & 3             &          \\[0.15cm]
15501                     & 05  &            & 00 56 08.6       & -72 35 02.5                &     -06-25 19:11               & 50.0     & 48.1 & 98           & 10            & 1 (1)    \\[0.15cm]
15498                     & 02  &            & 01 13 33.2       & -72 32 41.8                &     -08-09 00:04               & 50.0     & 49.4 & 87           & 9             &          \\
15500                     & 04  &            & 01 13 39.9       & -73 08 37.2                &     -08-24 21:07               & 50.0     & 49.4 & 67           & 8             &          \\
15502                     & 06  &            & 00 52 46.6       & -72 42 11.8                &     -08-26 11:44               & 50.0     & 45.8 & 63           & 4             &          \\
16320                     & 01  & 3          & 01 11 10.1       & -72 44 03.5                &     -08-30 03:28               & 32.0     & 29.7 & 59           & 1             &          \\
15497                     & 01  & 3          & 01 11 10.1       & -72 44 03.5                &     -08-31 16:27               & 18.0     & 17.8 & 43           & 1             &          \\[0.15cm]
15504                     & 08  &            & 00 56 35.9       & -72 20 06.3                &     -09-06 01:43               & 50.0     & 48.8 & 98           & 15            & 1        \\
15507                     & 11  & 4          & 00 47 09.6       & -73 07 33.0                &     -09-13 16:37               & 25.0     & 24.7 & 44           & 4             & 1 (1)    \\
16367                     & 11  & 4          & 00 47 09.6       & -73 07 33.0                &     -09-25 14:39               & 25.0     & 24.0 & 45           & 4             & 1 (1)    \\[0.15cm]
15499                     & 03  & 5          & 01 13 56.9       & -73 20 34.0                &     -10-14 22:57               & 25.0     & 24.0 & 46           & 4             &          \\
16490                     & 03  & 5          & 01 13 56.9       & -73 20 34.0                &     -10-15 20:36               & 25.0     & 24.2 & 46           & 2             &          \\[0.15cm]
15505                     & 09  &            & 00 49 43.2       & -72 49 16.3                & 2014-01-15 11:21               & 50.0     & 48.4 & 88           & 7             & 1        \\
15506                     & 10  &            & 00 41 00.0       & -73 20 00.0                &     -02-28 16:49               & 50.0     & 49.4 & 76           & 5             &          \\
15503                     & 07  &            & 00 51 52.0       & -73 00 24.6                &     -03-12 14:26               & 50.0     & 45.4 & 87           & 5             &          \\
\hline\multicolumn{6}{r}{Total} &	1400.0&	1344.3&	2339&	244&	18\\ \hline
\end{tabular*}

%% file: tab2.tex
\begin{tabular*}{\textwidth}{cc cc ccc}
\hline\hline
   (1)        & (2)              &       & (3)  & (4)          & (5)        & (6)      \\
DF            & Stacked          & Exp.  & GTI  & \multicolumn{3}{c}{Source Count} \\
\cline{5-7}ID & Obs.             &       &      & Total        & Net        & Periodic \\
              &                  & (ks)  & (ks) & 0.5--7 keV   &  $\ge$ 100 & Sources  \\
\hline
01A           & 7155, 7327       & 100.0 & 93.8 & 131          & 20         & 5        \\
02A           & 7156, 8479, 8481 & 100.0 & 96.7 & 118          & 19         & 4        \\
01            & 16320, 15497     & 50.0  & 47.5 & 78           & 4          &          \\
11            & 15497, 16320     & 50.0  & 48.7 & 66           & 4          & 1        \\
03            & 15499, 16490     & 50.0  & 48.2 & 65           & 5          &          \\
\hline\multicolumn{2}{r}{Total} &	350.0&	334.9&	458&	52&	10\\ \hline
\end{tabular*}

%% file: tab3.tex
\begin{tabular*}{\textwidth}{lc r@{}l D{.}{.}{3.1} r D{,}{}{3.1} r@{}l r@{}l r@{}l r@{}l ccccrcl}
\hline\hline
   (1)        & (2)                 & \multicolumn{2}{c}{(3)      } & \multicolumn{1}{c}{(4)}     & \multicolumn{1}{c}{(5)}   & \multicolumn{1}{c}{(6)}         & \multicolumn{2}{c}{(7) } & \multicolumn{2}{c}{(8)   } & \multicolumn{2}{c}{(9)       } & \multicolumn{2}{c}{(10)  } & (11)      & (12)        & (13)  & (14)     & (15) & (16) & (17)                     \\
Name          & Source              & \multicolumn{2}{c}{Pulsation} & \multicolumn{1}{c}{Orbital} & \multicolumn{1}{c}{LS}    & \multicolumn{1}{c}{False}       & \multicolumn{2}{c}{Mod.} & \multicolumn{2}{c}{Pulsed} & \multicolumn{2}{c}{Net       } & \multicolumn{2}{c}{Median} & Offset    & \multicolumn{2}{c}{Obs/Det/Pul} & Start    & GTIs & Flag & Obs.~IDs                 \\
\cline{17-18} & (CXOU J)            & \multicolumn{2}{c}{Period   } & \multicolumn{1}{c}{Period}  & \multicolumn{1}{c}{Power} & \multicolumn{1}{c}{Prob.}       & \multicolumn{2}{c}{Amp.} & \multicolumn{2}{c}{Frac. } & \multicolumn{2}{c}{Counts    } & \multicolumn{2}{c}{Energy} &           & Each        & Stack & Date     &      &      & (Deep Field ID)          \\
              &                     & \multicolumn{2}{c}{(s)      } & \multicolumn{1}{c}{(d)}     & \multicolumn{1}{c}{}      & \multicolumn{1}{c}{(10\sS{-x})} & \multicolumn{2}{c}{(\%)} & \multicolumn{2}{c}{(\%)  } & \multicolumn{2}{c}{0.3--8 keV} & \multicolumn{2}{c}{(keV) } & (\arcsec) &             &       & (Y/M/D)  & (ks) &      &                          \\
\hline
SXP7.92       & 005758.40--722229.5 & 7         & .918130(50)            & 35.61                       & 93.8                      & 33,.9                           & 76   & (4)  & 58     & (7)  & 2878       & (54)  & 3      & .29(5)  & 1.2       & 3/2/1       & 1/1/1 & 13/09/06 & 48   &      & 15504 (08)               \\
SXP18.3       & 004911.42--724936.9 & 18        & .36427(25)${}^\dagger$ & 17.79                       & 111.1                     & 41,.4                           & 54   & (6)  & 35     & (7)  & 4466       & (67)  & 3      & .49(3)  & 0.5       & 2/2/1       & 1/1/1 & 14/01/15 & 48   & a    & 15505 (09)               \\
SXP51.0       & 004814.15--731003.6 & 50        & .6684(17)${}^\dagger$  &                             & 148.4                     & 57,.6                           & 84   & (2)  & 74     & (4)  & 5947       & (77)  & 3      & .44(3)  & 0.3       & 3/1/1       & 2/1/1 & 13/03/04 & 45   &      & 14674 (11)               \\
SXP59.0       & 005456.18--722647.9 & 58        & .8345(22)${}^\dagger$  & 122                         & 40.2                      & 10,.6                           & 80   & (7)  & 67     & (12) & 398        & (20)  & 2      & .0(2)   & 0.7       & 3/3/1       & 2/2/1 & 06/04/25 & 93   & v    & 7155, 7327 (01A)         \\
SXP138        & 005323.89--722715.5 & 138       & .924(10)               & 125                         & 57.1                      & 17,.9                           & 72   & (8)  & 54     & (12) & 1357       & (37)  & 2      & .68(8)  & 0.1       & 2/2/2       & 1/1/1 & 06/04/25 & 93   &      & 7155, 7327 (01A)         \\[0.15cm]
SXP152        & 005750.41--720756.2 & 152       & .1063(94)${}^\dagger$  &                             & 87.6                      & 31,.2                           & 76   & (4)  & 68     & (5)  & 4883       & (70)  & 2      & .75(4)  & 0.5       & 1/1/1       & 0/0/0 & 01/05/15 & 98   & ei   & 1881                     \\
SXP172        & 005151.96--731033.9 & 171       & .848(20)               & 68.8                        & 33.4                      & 7,.6                            & 45   & (8)  & 25     & (9)  & 3263       & (57)  & 2      & .01(4)  & 0.5       & 3/3/1       & 1/1/1 & 06/11/21 & 96   & i    & 8479, 7156, 8481 (02A)   \\
SXP214        & 005011.26--730025.5 & 211       & .488(20)               &                             & 274.4                     & 112,.3                          & 96   & (3)  & 91     & (7)  & 1511       & (39)  & 3      & .84(5)  & 0.3       & 2/1/1       & 1/1/1 & 13/01/06 & 49   & ev   & 14670 (07)               \\
SXP304        & 010102.83--720659.2 & 304       & .542(66)               & 520                         & 28.3                      & 5,.4                            & 82   & (7)  & 66     & (14) & 319        & (19)  & 2      & .2(1)   & 0.8       & 1/1/1       & 0/0/0 & 01/05/15 & 98   &      & 1881                     \\
SXP323        & 005044.69--731605.2 & 317       & .292(54)               &                             & 51.9                      & 15,.7                           & 75   & (6)  & 64     & (8)  & 2377       & (49)  & 2      & .73(6)  & 0.2       & 3/3/2       & 1/1/1 & 06/11/21 & 96   & e    & 8479, 7156, 8481 (02A)   \\[0.15cm]
SXP327        & 005252.23--721715.3 & 326       & .813(31)${}^\dagger$   & 45.99                       & 193.0                     & 76,.9                           & 69   & (3)  & 56     & (4)  & 9071       & (96)  & 2      & .34(3)  & 0.6       & 2/2/2       & 1/1/1 & 06/04/25 & 93   & eia  & 7155, 7327 (01A)         \\
SXP504        & 005455.88--724511.0 & 501       & .29(27)                & 269                         & 53.7                      & 16,.4                           & 65   & (8)  & 39     & (13) & 1467       & (38)  & 3      & .04(8)  & 0.5       & 2/2/2       & 1/1/1 & 13/03/11 & 47   &      & 14669 (06)               \\
              & 005455.85--724511.5 & 501       & .35(20)${}^\dagger$    &                             & 105.7                     & 39,.0                           & 79   & (5)  & 63     & (9)  & 2060       & (46)  & 3      & .04(6)  & 1.1       & 2/2/2       & 1/1/1 & 13/08/26 & 45   &      & 15502 (06)               \\
SXP565        & 005735.98--721933.1 & 564       & .77(15)${}^\dagger$    & 152.4                       & 69.8                      & 23,.5                           & 67   & (7)  & 51     & (9)  & 2292       & (50)  & 1      & .94(3)  & 0.7       & 3/3/1       & 1/1/0 & 01/05/15 & 98   & ia   & 1881                     \\
SXP701        & 005518.44--723851.8 & 726       & .31(44)                & 412                         & 81.4                      & 28,.5                           & 88   & (5)  & 76     & (11) & 916        & (30)  & 2      & .96(6)  & 0.5       & 2/2/2       & 1/1/1 & 13/01/02 & 49   &      & 14668 (05)               \\
              & 005518.48--723852.4 & 728       & .11(57)                &                             & 52.6                      & 16,.0                           & 75   & (7)  & 57     & (13) & 515        & (23)  & 2      & .8(1)   & 1.1       & 2/2/2       & 1/1/1 & 13/06/25 & 48   &      & 15501 (05)               \\
SXP756        & 004942.02--732314.5 & 746       & .43(22)${}^\dagger$    & 394                         & 100.7                     & 36,.9                           & 63   & (5)  & 38     & (7)  & 4194       & (65)  & 2      & .86(4)  & 0.1       & 4/3/3       & 1/1/1 & 06/11/21 & 96   & a    & 8479, 7156, 8481 (02A)   \\[0.15cm]
SXP893        & 004929.80--731058.4 & 894       & .07(38)${}^\dagger$    &                             & 67.9                      & 22,.6                           & 85   & (6)  & 77     & (9)  & 1464       & (39)  & 2      & .99(7)  & 0.4       & 6/6/5       & 3/3/3 & 06/11/21 & 96   & e    & 8479, 7156, 8481 (02A)   \\
              & 004929.63--731057.7 & 892       & .55(76)                &                             & 65.4                      & 21,.6                           & 81   & (6)  & 67     & (11) & 1027       & (32)  & 3      & .16(9)  & 1.1       & 6/6/5       & 3/3/3 & 13/09/13 & 48   & e    & 15507, 16367 (11)        \\
\hline \multicolumn{7}{l}{\it Marginal or Unusual Detection}  \\
SXP5.05       & 005702.12--722555.9 & 961       & .6(1.4)                & 17.2                        & 25.0                      & 4,.0                            & 18   & (3)  & 9      & (3)  & 34179      & (185) & 3      & .39(1)  & 1.1       & 4/1/1       & 2/1/0 & 13/09/06 & 48   & v    & 15504 (08)               \\
              &                     & 3021      & (13)                   &                             & 28.1                      & 5,.4                            & 19   & (3)  & 10     & (3)  &            &       &        &         &           &             &       &          &      &      &                          \\
SXP7.59?      & 003942.37--732427.4 & 7         & .58712(12)             &                             & 17.5                      & 17.5                            & 58   & (11) & 38     & (17) & 151        & (12)  & 3      & .0(1)   & -         & 2/2/1       & 0/0/0 & 13/01/18 & 44   &      & 14673 (10)               \\
SXP8.02       & 010043.07--721133.4 & 8         & .018944(58)            &                             & 18.0                      & 10.5                            & 31   & (7)  & 17     & (7)  & 5592       & (75)  & 1      & .349(7) & 1.9       & 1/1/1       & 0/0/0 & 01/05/15 & 98   &      & 1881                     \\
SXP74.7       & 004903.35--725052.2 & 24        & .9452(12)              & 33.3                        & 18.9                      & 4.5                             & 56   & (12) & 38     & (15) & 912        & (30)  & 2      & .62(9)  & 0.2       & 2/2/1       & 1/1/1 & 13/01/17 & 44   &      & 14672 (09)               \\
              &                     & 74        & .836(23)${}^\dagger$   &                             & 4.1                       & -                               & 59   & (11) & 39     & (14) &            &       &        &         &           &             &       &          &      &      &                          \\
SXP4693       & 005446.38--722522.9 & 4685      & (22)                   &                             & 16.5                      & 48.6                            & 68   & (9)  & 49     & (16) & 151        & (13)  & 2      & .3(2)   & 0.3       & 3/2/0       & 2/1/0 & 06/04/25 & 93   &      & 7155, 7327 (01A)         \\
\hline \multicolumn{5}{l}{\it No Pulsation Detection} \\
SXP7.78       & 005205.88--722604.9 &           &                        & 44.8                        &                           &                                 & $<$  & 100  &        &      & 73         & (9)   & 2      & .1(3)   & 1.3       & 2/2/0       & 1/1/0 & 06/04/25 & 93   &      & 7155, 7327 (01A)         \\
SXP8.88       & 005153.05--723149.1 &           &                        &                             &                           &                                 & $<$  & 67   &        &      & 182        & (15)  & 2      & .3(2)   & 1.0       & 2/2/0       & 1/1/0 & 06/04/25 & 93   &      & 7155, 7327 (01A)         \\
SXP9.13       & 004913.53--731138.1 &           &                        & 40.1                        &                           &                                 & $<$  & 26   &        &      & 1178       & (35)  & 2      & .91(7)  & 1.2       & 6/6/0       & 3/3/0 & 06/11/21 & 96   &      & 8479, 7156, 8481 (02A)   \\
              & 004913.51--731137.4 &           &                        &                             &                           &                                 & $<$  & 33   &        &      & 761        & (28)  & 2      & .97(9)  & 0.6       & 6/6/0       & 3/3/0 & 13/03/04 & 94   &      & 15507, 16367, 14674 (11) \\
SXP15.3       & 005213.93--731918.4 &           &                        & 74.3                        &                           &                                 & $<$  & 62   &        &      & 213        & (15)  & 2      & .3(2)   & 0.6       & 3/3/0       & 1/1/0 & 06/11/21 & 96   &      & 8479, 7156, 8481 (02A)   \\
SXP46.6       & 005355.37--722645.5 &           &                        & 137.4                       &                           &                                 & $<$  & 100  &        &      & 64         & (8)   & 1      & .6(1)   & 0.3       & 2/2/0       & 1/1/0 & 06/04/25 & 93   &      & 7155, 7327 (01A)         \\[0.15cm]
SXP82.4       & 005209.01--723803.5 &           &                        & 362                         &                           &                                 & $<$  & 100  &        &      & 18         & (5)   & 1      & .9(4)   & 0.5       & 5/2/0       & 1/1/0 & 13/08/26 & 45   &      & 15502 (06)               \\
SXP140        & 005605.42--722200.0 &           &                        & 197                         &                           &                                 & $<$  & 100  &        &      & 12         & (5)   & 2      & (1)     & 1.2       & 3/2/0       & 1/1/0 & 13/09/06 & 48   &      & 15504 (08)               \\
SXP264        & 004723.40--731227.1 &           &                        & 49.2                        &                           &                                 & $<$  & 28   &        &      & 1057       & (33)  & 2      & .70(8)  & 0.1       & 3/3/0       & 2/2/0 & 13/03/04 & 94   &      & 15507, 16367, 14674 (11) \\
SXP292        & 005047.98--731817.9 &           &                        &                             &                           &                                 & $<$  & 100  &        &      & 20         & (5)   & 2      & .1(3)   & 0.2       & 3/2/0       & 1/1/0 & 06/11/21 & 42   &      & 8479 (02A)               \\
SXP342        & 005403.87--722632.9 &           &                        &                             &                           &                                 & $<$  & 100  &        &      & 17         & (5)   & 3      & .7(7)   & 0.2       & 2/2/0       & 1/1/0 & 06/04/26 & 47   &      & 7327 (01A)               \\[0.15cm]
SXP645        & 005535.14--722906.7 &           &                        &                             &                           &                                 & $<$  & 48   &        &      & 355        & (19)  & 2      & .80(9)  & 0.7       & 3/3/0       & 2/2/0 & 13/01/02 & 97   &      & 14668, 15501 (05)        \\
\hline
\end{tabular*}

%% file: tab4.tex
\begin{tabular*}{\textwidth}{l r@{}l r@{}l r@{}l r@{}l rr r@{}l r@{}l r@{\ }l r@{}l r@{}l r@{}l r@{}l l}
\hline\hline
   (1)                                                                      & \multicolumn{2}{c}{(2)       } & \multicolumn{2}{c}{(3)     } & \multicolumn{2}{c}{(4)              } & \multicolumn{2}{c}{(5)     } & (6)               & (7)           & \multicolumn{2}{c}{(8)             } & \multicolumn{2}{c}{(9)     } & \multicolumn{2}{c}{(10)            } & \multicolumn{2}{c}{(11)                                        } & \multicolumn{2}{c}{(12)    } & \multicolumn{2}{c}{(13)                               } & \multicolumn{2}{c}{(14)     } & (15)                     \\
Name                                                                        & \multicolumn{2}{c}{Net       } & \multicolumn{2}{c}{Hardness} & \multicolumn{6}{c}{Quantile Analysis} & \multicolumn{6}{c}{Spectral Fit    } & \multicolumn{4}{c}{$F$\Ss{X}                                   } & \multicolumn{4}{c}{$L$\Ss{X}                          } & Obs.~IDs                 \\
\cmidrule(r){6-11}\cmidrule(r){12-17}\cmidrule(r){18-21}\cmidrule(r){22-25} & \multicolumn{2}{c}{Counts    } & \multicolumn{2}{c}{Ratio   } & \multicolumn{2}{c}{Median           } & \multicolumn{2}{c}{Quartile} & $\Gamma$          & nH            & \multicolumn{2}{c}{$\Gamma$        } & \multicolumn{2}{c}{nH      } & \multicolumn{2}{c}{$\chi^2_r$ / DoF} & \multicolumn{2}{c}{0.5--2 keV                                  } & \multicolumn{2}{c}{2--8 keV} & \multicolumn{2}{c}{Observed                           } & \multicolumn{2}{c}{Intrinsic} & (Deep Field ID)          \\
                                                                            & \multicolumn{2}{c}{0.3--8 keV} & \multicolumn{2}{c}{        } & \multicolumn{2}{c}{(keV)            } & \multicolumn{2}{c}{Ratio   } &  or $kT$$\dagger$ &               & \multicolumn{2}{c}{ or  $kT$$\ddag$} & \multicolumn{2}{c}{        } & \multicolumn{2}{c}{                } & \multicolumn{4}{c}{($\times$ 10\sS{-13} erg cm\sS{-2} s\sS{-1})} & \multicolumn{4}{c}{0.5--8 keV (10\sS{34} erg s\sS{-1})} &                          \\
\hline
SXP7.92                                                                     & 390        & (20)  & 0        & .52(5)  & 3                 & .4(1)   & 1        & .11(5)  & 0.37              & 0.63          &                  &        &          &         &                  &       & 0                                            & .1(1)  & 2        & (2)    & 10                                  & .0(5)  & 10        & .4(5)  & 14671 (08)               \\
                                                                            & 2878       & (54)  & 0        & .42(2)  & 3                 & .29(5)  & 1        & .05(2)  & 0.35              & 0.33          & 0                & .25(6) & 0        & .1(2)   & 0.86             & / 60  & 1                                            & .3(1)  & 16       & (1)    & 75                                  & (6)    & 76        & (7)    & 15504 (08)               \\
SXP18.3                                                                     & 189        & (14)  & 0        & .4(1)   & 3                 & .2(2)   & 1        & .14(7)  & 0.84              & 2.48          &                  &        &          &         &                  &       & 0                                            & .08(8) & 0        & .8(8)  & 3                                   & .8(3)  & 4         & .4(3)  & 14672 (09)               \\
                                                                            & 4466       & (67)  & 0        & .57(1)  & 3                 & .49(3)  & 1        & .24(2)  & 0.62              & 2.61          & 0                & .53(6) & 2        & .5(3)   & 1.12             & / 94  & 1                                            & .3(1)  & 23       & (2)    & 103                                 & (9)    & 110       & (10)   & 15505 (09)               \\
SXP51.0                                                                     & 5947       & (77)  & 0        & .54(1)  & 3                 & .44(3)  & 1        & .21(2)  & 0.56              & 2.23          & 0                & .50(5) & 2        & .3(2)   & 0.87             & / 98  & 2                                            & .1(1)  & 37       & (3)    & 170                                 & (12)   & 190       & (14)   & 14674 (11)               \\
SXP59.0                                                                     & 398        & (20)  & 0        & .03(5)  & 2                 & .0(2)   & 0        & .94(4)  & 0.98              & 0.10          & 0                & .9(2)  & 0        & .3(5)   & 1.32             & / 14  & 0                                            & .12(3) & 0        & .7(2)  & 4                                   & (1)    & 4         & (1)    & 7155, 7327 (01A)         \\
                                                                            & 67         & (9)   & 0        & .1(1)   & 2                 & .2(3)   & 1        & .3(2)   & 2.55              & 5.53          &                  &        &          &         &                  &       & 0                                            & .05(5) & 0        & .2(2)  & 1                                   & .0(1)  & 2         & .9(4)  & 14671 (08)               \\
SXP138                                                                      & 1357       & (37)  & 0        & .23(3)  & 2                 & .68(8)  & 0        & .98(3)  & 0.76              & 0.60          & 0                & .70(9) & 0        & .3(2)   & 0.95             & / 37  & 0                                            & .64(8) & 5        & .0(6)  & 24                                  & (3)    & 25        & (3)    & 7155, 7327 (01A)         \\[0.15cm]
SXP152                                                                      & 4883       & (70)  & 0        & .25(1)  & 2                 & .75(4)  & 1        & .03(1)  & 0.85              & 1.28          & 0                & .76(4) & 0        & .9(1)   & 1.02             & / 76  & 1                                            & .16(7) & 10       & .8(6)  & 52                                  & (3)    & 56        & (3)    & 1881                     \\
SXP172                                                                      & 3263       & (57)  & 0        & .01(2)  & 2                 & .01(4)  & 0        & .93(2)  & 1.10              & 0.00          & 1                & .08(3) & 0        & .00(2)  & 1.30             & / 52  & 1                                            & .69(6) & 6        & .7(2)  & 36                                  & (1)    & 37        & (1)    & 8479, 7156, 8481 (02A)   \\
SXP214                                                                      & 1511       & (39)  & 0        & .67(3)  & 3                 & .84(5)  & 1        & .46(4)  & 0.19              & 2.77          & 0                & .0(1)  & 2        & .1(7)   & 1.00             & / 31  & 0                                            & .35(7) & 11       & (2)    & 47                                  & (9)    & 50        & (9)    & 14670 (07)               \\
SXP304                                                                      & 319        & (19)  & 0        & .08(7)  & 2                 & .2(1)   & 0        & .96(5)  & 0.95              & 0.37          & 0                & .9(2)  & 0        & .2(6)   & 0.58             & / 11  & 0                                            & .10(3) & 0        & .6(2)  & 3                                   & .0(9)  & 3         & (1)    & 1881                     \\
SXP323                                                                      & 2377       & (49)  & 0        & .26(2)  & 2                 & .73(6)  & 1        & .04(2)  & 0.97              & 1.49          & 0                & .87(7) & 1        & .0(2)   & 1.51             & / 39  & 0                                            & .51(5) & 4        & .3(4)  & 21                                  & (2)    & 23        & (2)    & 8479, 7156, 8481 (02A)   \\[0.15cm]
SXP327                                                                      & 9071       & (96)  & 0        & .12(1)  & 2                 & .34(3)  & 1        & .02(1)  & 1.00              & 0.69          & 0                & .90(3) & 0        & .37(7)  & 1.38             & / 70  & 3                                            & .2(1)  & 20       & .1(8)  & 100                                 & (4)    & 107       & (4)    & 7155, 7327 (01A)         \\
SXP504                                                                      & 1467       & (38)  & 0        & .38(3)  & 3                 & .04(8)  & 1        & .12(2)  & 0.81              & 1.13          & 0                & .8(1)  & 1        & .3(4)   & 0.65             & / 40  & 0                                            & .8(1)  & 8        & (1)    & 39                                  & (6)    & 43        & (7)    & 14669 (06)               \\
                                                                            & 2060       & (46)  & 0        & .41(3)  & 3                 & .04(6)  & 1        & .10(2)  & 0.76              & 0.63          & 0                & .60(8) & 0        & .3(3)   & 0.91             & / 43  & 1                                            & .4(2)  & 13       & (1)    & 61                                  & (7)    & 63        & (7)    & 15502 (06)               \\
SXP565                                                                      & 2292       & (50)  & -0       & .04(3)  & 1                 & .94(3)  & 1        & .15(3)  & 1.40              & 1.59          & 1                & .16(7) & 1        & .0(1)   & 1.06             & / 47  & 0                                            & .63(5) & 3        & .8(3)  & 19                                  & (2)    & 22        & (2)    & 1881                     \\
                                                                            & 2147       & (46)  & 0        & .39(2)  & 2                 & .97(5)  & 1        & .16(2)  & 1.12              & 3.17          & 1                & .07(9) & 2        & .8(4)   & 0.78             & / 35  & 0                                            & .48(6) & 5        & .0(7)  & 24                                  & (3)    & 29        & (4)    & 14671, 15504 (08)        \\
SXP701                                                                      & 916        & (30)  & 0        & .38(3)  & 2                 & .96(6)  & 1        & .16(4)  & 1.21              & 3.33          & 0                & .9(1)  & 2        & .2(6)   & 1.03             & / 38  & 0                                            & .4(1)  & 5        & (1)    & 22                                  & (5)    & 26        & (6)    & 14668 (05)               \\
                                                                            & 515        & (23)  & 0        & .30(6)  & 2                 & .8(1)   & 1        & .08(4)  & 1.02              & 1.63          & 1                & .0(2)  & 1        & .7(8)   & 0.59             & / 20  & 0                                            & .26(8) & 2        & .2(6)  & 10                                  & (3)    & 12        & (4)    & 15501 (05)               \\
SXP756                                                                      & 4194       & (65)  & 0        & .30(2)  & 2                 & .86(4)  & 1        & .06(2)  & 0.76              & 1.19          & 0                & .76(5) & 1        & .1(2)   & 0.60             & / 71  & 1                                            & .02(7) & 10       & .0(7)  & 47                                  & (3)    & 52        & (4)    & 8479, 7156, 8481 (02A)   \\[0.15cm]
SXP893                                                                      & 1464       & (39)  & 0        & .39(3)  & 2                 & .99(7)  & 1        & .10(2)  & 0.76              & 1.74          & 0                & .7(1)  & 1        & .6(5)   & 0.63             & / 23  & 0                                            & .32(5) & 3        & .6(6)  & 17                                  & (3)    & 19        & (3)    & 8479, 7156, 8481 (02A)   \\
                                                                            & 509        & (23)  & 0        & .52(5)  & 3                 & .18(8)  & 1        & .27(6)  & 1.09              & 3.39          &                  &        &          &         &                  &       & 0                                            & .6(6)  & 6        & (6)    & 30                                  & (1)    & 37        & (2)    & 14674 (11)               \\
                                                                            & 1027       & (32)  & 0        & .42(4)  & 3                 & .16(9)  & 1        & .12(3)  & 0.54              & 0.84          & 0                & .6(1)  & 0        & .7(6)   & 0.68             & / 28  & 0                                            & .6(1)  & 6        & (1)    & 29                                  & (6)    & 30        & (6)    & 15507, 16367 (11)        \\
\hline \multicolumn{7}{l}{\it Marginal or Unusual Detection}  \\
SXP5.05                                                                     & 34179      & (185) & 0        & .441(5) & 3                 & .39(1)  & 1        & .041(6) & 0.20              & 0.00          & 0                & .17(1) & 0        & .000(4) & 1.37             & / 145 & 14                                           & .61(8) & 197      & (1)    & 911                                 & (5)    & 918       & (5)    & 15504 (08)               \\
SXP7.59                                                                     & 151        & (12)  & 0        & .47(9)  & 3                 & .0(1)   & 1        & .3(1)   & 1.74              & 7.01          &                  &        &          &         &                  &       & 0                                            & .06(6) & 0        & .7(7)  & 3                                   & .2(3)  & 5         & .8(5)  & 14673 (10)               \\
                                                                            & 308        & (18)  & 0        & .08(6)  & 2                 & .1(1)   & 1        & .14(6)  & 1.73              & 1.99          & 1                & .4(4)  & 1        & (1)     & 0.60             & / 5   & 0                                            & .3(2)  & 1        & .3(8)  & 7                                   & (4)    & 8         & (5)    & 15506 (10)               \\
SXP8.02                                                                     & 5592       & (75)  & -0       & .72(1)  & 1                 & .349(7) & 1        & .63(2)  & $\dagger$1.13     & $\dagger$0.72 & $\ddag$0         & .39(6) & $\ddag$0 & .0(1)   & 1.15             & / 43  & $\ddag$3                                     & .01(7) & $\ddag$1 & .67(4) & $\ddag$20                           & .1(5)  & $\ddag$21 & .4(5)  & 1881                     \\
                                                                            &            &       &          &         &                   &         &          &         &                   &               & $\ddag$3         & .7(6)  &          &         &                  &       &                                              &        &          &        &                                     &        &           &        &                          \\
SXP74.7                                                                     & 912        & (30)  & 0        & .22(4)  & 2                 & .62(9)  & 1        & .04(3)  & 1.09              & 1.25          & 1                & .0(1)  & 1        & .1(5)   & 0.55             & / 24  & 0                                            & .50(9) & 3        & .5(7)  & 17                                  & (3)    & 20        & (4)    & 14672 (09)               \\
                                                                            & 196        & (14)  & 0        & .25(8)  & 2                 & .5(2)   & 1        & .19(9)  & 1.81              & 4.03          &                  &        &          &         &                  &       & 0                                            & .09(9) & 0        & .6(6)  & 3                                   & .0(2)  & 5         & .0(4)  & 15505 (09)               \\
SXP4693                                                                     & 151        & (13)  & 0        & .09(9)  & 2                 & .3(2)   & 1        & .25(9)  & 2.08              & 4.51          &                  &        &          &         &                  &       & 0                                            & .05(5) & 0        & .2(2)  & 1                                   & .2(1)  & 2         & .5(2)  & 7155, 7327 (01A)         \\
\hline \multicolumn{5}{l}{\it No Pulsation Detection} \\
SXP7.78                                                                     & 73         & (9)   & 0        & .1(2)   & 2                 & .1(3)   & 0        & .9(1)   & 0.80              & 0.00          &                  &        &          &         &                  &       & 0                                            & .03(3) & 0        & .2(2)  & 0                                   & .9(1)  & 0         & .9(1)  & 7155, 7327 (01A)         \\
SXP8.88                                                                     & 182        & (15)  & 0        & .1(1)   & 2                 & .3(2)   & 1        & .08(8)  & 1.28              & 1.27          &                  &        &          &         &                  &       & 0                                            & .07(7) & 0        & .4(4)  & 2                                   & .0(2)  & 2         & .4(2)  & 7155, 7327 (01A)         \\
SXP9.13                                                                     & 1178       & (35)  & 0        & .37(3)  & 2                 & .91(7)  & 1        & .13(3)  & 1.01              & 2.56          & 1                & .1(1)  & 2        & .9(5)   & 0.83             & / 24  & 0                                            & .28(5) & 3        & .0(5)  & 14                                  & (2)    & 17        & (3)    & 8479, 7156, 8481 (02A)   \\
                                                                            & 761        & (28)  & 0        & .41(4)  & 2                 & .97(9)  & 1        & .18(3)  & 1.09              & 2.82          & 0                & .9(2)  & 1        & .9(8)   & 0.87             & / 24  & 0                                            & .30(8) & 2        & .9(8)  & 14                                  & (4)    & 16        & (4)    & 15507, 16367, 14674 (11) \\
SXP15.3                                                                     & 213        & (15)  & 0        & .16(7)  & 2                 & .3(2)   & 1        & .01(6)  & 1.00              & 0.75          & 0                & .8(4)  & 0        & (2)     & 1.19             & / 6   & 0                                            & .09(7) & 0        & .6(4)  & 3                                   & (2)    & 3         & (2)    & 8479, 7156, 8481 (02A)   \\
SXP46.6                                                                     & 64         & (8)   & -0       & .5(1)   & 1                 & .6(1)   & 1        & .3(2)   & 2.19              & 1.01          &                  &        &          &         &                  &       & 0                                            & .03(3) & 0        & .04(4) & 0                                   & .32(4) & 0         & .49(6) & 7155, 7327 (01A)         \\[0.15cm]
SXP82.4                                                                     & 18         & (5)   & -0       & .1(2)   & 1                 & .9(4)   & 0        & .9(2)   & 1.10              & 0.00          &                  &        &          &         &                  &       & 0                                            & .02(2) & 0        & .06(6) & 0                                   & .4(1)  & 0         & .4(1)  & 15502 (06)               \\
SXP140                                                                      & 12         & (5)   & -0       & .2(3)   & 2                 & (1)     & 0        & .5(3)   & 1.40              & 0.00          &                  &        &          &         &                  &       & 0                                            & .01(1) & 0        & .03(3) & 0                                   & .18(7) & 0         & .19(7) & 15504 (08)               \\
SXP264                                                                      & 1057       & (33)  & 0        & .24(3)  & 2                 & .70(8)  & 1        & .08(3)  & 1.04              & 1.62          & 1                & .0(1)  & 1        & .2(4)   & 0.75             & / 35  & 0                                            & .32(5) & 2        & .4(4)  & 11                                  & (2)    & 13        & (2)    & 15507, 16367, 14674 (11) \\
SXP292                                                                      & 20         & (5)   & 0        & .2(3)   & 2                 & .1(3)   & 1        & .6(4)   & 4.08              & 12.02         &                  &        &          &         &                  &       & 0                                            & .01(1) & 0        & .05(5) & 0                                   & .26(7) & 6         & (2)    & 8479 (02A)               \\
SXP342                                                                      & 17         & (5)   & 0        & .6(2)   & 3                 & .7(7)   & 1        & .1(3)   & -0.10             & 0.00          &                  &        &          &         &                  &       & 0                                            & .01(1) & 0        & .2(2)  & 0                                   & .8(3)  & 0         & .8(3)  & 7327 (01A)               \\[0.15cm]
SXP645                                                                      & 355        & (19)  & 0        & .29(6)  & 2                 & .80(9)  & 1        & .06(5)  & 0.85              & 1.11          & 0                & .5(2)  & 0        & .0(9)   & 1.10             & / 9   & 0                                            & .14(4) & 1        & .1(4)  & 5                                   & (2)    & 5         & (2)    & 14668, 15501 (05)        \\
\hline
\end{tabular*}

%% file: tab5.tex
\begin{tabular*}{\textwidth}{c c r@{}l r@{}l  r@{\ }l r@{}l r@{}l}
\hline\hline
   \multicolumn{1}{c}{(1)}                     & (2)   & \multicolumn{2}{c}{(3)     } & \multicolumn{2}{c}{(4)} & \multicolumn{2}{c}{(5)             } & \multicolumn{2}{c}{(6)      } & \multicolumn{2}{c}{(7) } \\
Data                                           & Net   & \multicolumn{2}{c}{$\Gamma$} & \multicolumn{2}{c}{\nH} & \multicolumn{2}{c}{$\chi^2_r$ / DoF} & \multicolumn{4}{c}{$L$\Ss{X}} \\
\cmidrule(r){9-12}Segment                      & Count & \multicolumn{2}{c}{        } & \multicolumn{2}{c}{   } & \multicolumn{2}{c}{                } & \multicolumn{2}{c}{Obs.     } & \multicolumn{2}{c}{Int.} \\
\hline
\multicolumn{1}{l}{\it All}                    & 4466  & $0.5$    & $_{-0.1}^{+0.1}$ & 2.5 & $_{-0.3}^{+0.3}$ & 1.12             & / 94 & 10.3      & $_{-0.9}^{+0.9}$ & 11.5 & $_{-1.0}^{+1.0}$ \\
\hline\multicolumn{3}{l}{\it By folded phases} &                  &     &                  &                  &      &           &                  &      &                  \\
P0.15--0.35                                    & 1215  & $0.5$    & $_{-0.1}^{+0.1}$ & 3.2 & $_{-0.8}^{+0.9}$ & 1.21             & / 34 & 14.7      & $_{-2.7}^{+3.2}$ & 16.4 & $_{-3.0}^{+3.6}$ \\
P0.70--1.00                                    & 1060  & $0.7$    & $_{-0.1}^{+0.1}$ & 2.3 & $_{-0.6}^{+0.6}$ & 0.88             & / 45 & 7.6       & $_{-1.3}^{+1.5}$ & 8.5  & $_{-1.4}^{+1.7}$ \\
\hline
\end{tabular*}

%% file: tab6.tex
\begin{tabular*}{\textwidth}{c c r@{}l r@{}l  r@{\ }l r@{}l r@{}l}
\hline\hline
   \multicolumn{1}{c}{(1)}                     & (2)   & \multicolumn{2}{c}{(3)     } & \multicolumn{2}{c}{(4)} & \multicolumn{2}{c}{(5)             } & \multicolumn{2}{c}{(6)      } & \multicolumn{2}{c}{(7) } \\
Data                                           & Net   & \multicolumn{2}{c}{$\Gamma$} & \multicolumn{2}{c}{\nH} & \multicolumn{2}{c}{$\chi^2_r$ / DoF} & \multicolumn{4}{c}{$L$\Ss{X}} \\
\cmidrule(r){9-12}Segment                      & Count & \multicolumn{2}{c}{        } & \multicolumn{2}{c}{   } & \multicolumn{2}{c}{                } & \multicolumn{2}{c}{Obs.     } & \multicolumn{2}{c}{Int.} \\
\hline
\multicolumn{1}{l}{\it All}                    & 6088  & $0.5$    & $_{-0.0}^{+0.0}$ & 2.3 & $_{-0.2}^{+0.2}$ & 0.87             & / 98 & 16.9      & $_{-1.1}^{+1.2}$ & 18.7 & $_{-1.3}^{+1.4}$ \\
\hline\multicolumn{3}{l}{\it By folded phases} &                  &     &                  &                  &      &           &                  &      &                  \\
P0.02--0.18                                    & 729   & $0.9$    & $_{-0.2}^{+0.2}$ & 3.1 & $_{-0.8}^{+0.8}$ & 1.01             & / 30 & 10.8      & $_{-2.2}^{+2.8}$ & 12.9 & $_{-2.7}^{+3.3}$ \\
P0.18--0.47                                    & 1768  & $0.4$    & $_{-0.1}^{+0.1}$ & 2.3 & $_{-0.5}^{+0.6}$ & 0.77             & / 49 & 17.3      & $_{-2.4}^{+2.8}$ & 19.0 & $_{-2.6}^{+3.0}$ \\
P0.47--0.60                                    & 254   & $0.3$    & $_{-0.3}^{+0.4}$ & 0.8 & $_{-0.0}^{+2.6}$ & 0.93             & / 8  & 5.4       & $_{-1.7}^{+4.2}$ & 5.6  & $_{-1.7}^{+4.4}$ \\
P0.60--0.02                                    & 3337  & $0.5$    & $_{-0.1}^{+0.1}$ & 2.4 & $_{-0.3}^{+0.3}$ & 1.07             & / 69 & 22.2      & $_{-2.1}^{+2.3}$ & 24.5 & $_{-2.3}^{+2.6}$ \\
\hline
\end{tabular*}

%% file: tab7.tex
\begin{tabular*}{\textwidth}{c c r@{}l r@{}l  r@{\ }l r@{}l r@{}l}
\hline\hline
   \multicolumn{1}{c}{(1)}                     & (2)   & \multicolumn{2}{c}{(3)     } & \multicolumn{2}{c}{(4)} & \multicolumn{2}{c}{(5)             } & \multicolumn{2}{c}{(6)      } & \multicolumn{2}{c}{(7) } \\
Data                                           & Net   & \multicolumn{2}{c}{$\Gamma$} & \multicolumn{2}{c}{\nH} & \multicolumn{2}{c}{$\chi^2_r$ / DoF} & \multicolumn{4}{c}{$L$\Ss{X}} \\
\cmidrule(r){9-12}Segment                      & Count & \multicolumn{2}{c}{        } & \multicolumn{2}{c}{   } & \multicolumn{2}{c}{                } & \multicolumn{2}{c}{Obs.     } & \multicolumn{2}{c}{Int.} \\
\hline
\multicolumn{1}{l}{\it All}                    & 4883  & $0.8$    & $_{-0.0}^{+0.0}$ & 0.9 & $_{-0.1}^{+0.1}$ & 1.02             & / 76 & 5.2       & $_{-0.3}^{+0.3}$ & 5.6  & $_{-0.3}^{+0.3}$ \\
\hline\multicolumn{3}{l}{\it By folded phases} &                  &     &                  &                  &      &           &                  &      &                  \\
P0.10--0.65                                    & 3174  & $0.7$    & $_{-0.1}^{+0.1}$ & 0.9 & $_{-0.1}^{+0.2}$ & 0.91             & / 66 & 6.2       & $_{-0.4}^{+0.5}$ & 6.7  & $_{-0.5}^{+0.5}$ \\
P0.75--0.90                                    & 587   & $1.0$    & $_{-0.1}^{+0.2}$ & 1.2 & $_{-0.5}^{+0.5}$ & 1.17             & / 23 & 3.5       & $_{-0.6}^{+0.8}$ & 4.0  & $_{-0.7}^{+0.9}$ \\
\hline
\end{tabular*}

%% file: tab8.tex
\begin{tabular*}{\textwidth}{c c r@{}l r@{}l  r@{\ }l r@{}l r@{}l}
\hline\hline
   \multicolumn{1}{c}{(1)}                     & (2)   & \multicolumn{2}{c}{(3)     } & \multicolumn{2}{c}{(4)} & \multicolumn{2}{c}{(5)             } & \multicolumn{2}{c}{(6)      } & \multicolumn{2}{c}{(7) } \\
Data                                           & Net   & \multicolumn{2}{c}{$\Gamma$} & \multicolumn{2}{c}{\nH} & \multicolumn{2}{c}{$\chi^2_r$ / DoF} & \multicolumn{4}{c}{$L$\Ss{X}} \\
\cmidrule(r){9-12}Segment                      & Count & \multicolumn{2}{c}{        } & \multicolumn{2}{c}{   } & \multicolumn{2}{c}{                } & \multicolumn{2}{c}{Obs.     } & \multicolumn{2}{c}{Int.} \\
\hline
\multicolumn{1}{l}{\it All}                    & 3289  & $1.1$    & $_{-0.0}^{+0.0}$ & 0.0 & $_{-0.0}^{+0.0}$ & 1.30             & / 52 & 3.6       & $_{-0.1}^{+0.1}$ & 3.7  & $_{-0.1}^{+0.1}$ \\
\hline\multicolumn{3}{l}{\it By folded phases} &                  &     &                  &                  &      &           &                  &      &                  \\
P0.30--0.70                                    & 1529  & $0.8$    & $_{-0.0}^{+0.0}$ & 0.0 & $_{-0.0}^{+0.0}$ & 1.02             & / 64 & 4.6       & $_{-0.2}^{+0.2}$ & 4.7  & $_{-0.2}^{+0.2}$ \\
P0.75--0.25                                    & 1429  & $1.4$    & $_{-0.1}^{+0.1}$ & 0.0 & $_{-0.0}^{+0.1}$ & 0.82             & / 53 & 2.7       & $_{-0.1}^{+0.2}$ & 2.8  & $_{-0.1}^{+0.2}$ \\
\hline
\end{tabular*}

%% file: tab9.tex
\begin{tabular*}{\textwidth}{c c r@{}l r@{}l  r@{\ }l r@{}l r@{}l}
\hline\hline
   \multicolumn{1}{c}{(1)}                     & (2)   & \multicolumn{2}{c}{(3)     } & \multicolumn{2}{c}{(4)} & \multicolumn{2}{c}{(5)             } & \multicolumn{2}{c}{(6)      } & \multicolumn{2}{c}{(7) } \\
Data                                           & Net   & \multicolumn{2}{c}{$\Gamma$} & \multicolumn{2}{c}{\nH} & \multicolumn{2}{c}{$\chi^2_r$ / DoF} & \multicolumn{4}{c}{$L$\Ss{X}} \\
\cmidrule(r){9-12}Segment                      & Count & \multicolumn{2}{c}{        } & \multicolumn{2}{c}{   } & \multicolumn{2}{c}{                } & \multicolumn{2}{c}{Obs.     } & \multicolumn{2}{c}{Int.} \\
\hline
\multicolumn{1}{l}{\it All}                    & 9139  & $0.9$    & $_{-0.0}^{+0.0}$ & 0.4 & $_{-0.1}^{+0.1}$ & 1.38             & / 70 & 10.0      & $_{-0.4}^{+0.4}$ & 10.7 & $_{-0.4}^{+0.4}$ \\
\hline\multicolumn{3}{l}{\it By folded phases} &                  &     &                  &                  &      &           &                  &      &                  \\
P0.05--0.30                                    & 2843  & $1.0$    & $_{-0.1}^{+0.1}$ & 0.3 & $_{-0.1}^{+0.1}$ & 1.06             & / 61 & 11.8      & $_{-0.8}^{+0.9}$ & 12.6 & $_{-0.8}^{+0.9}$ \\
P0.30--0.65                                    & 2996  & $0.7$    & $_{-0.1}^{+0.1}$ & 0.1 & $_{-0.0}^{+0.1}$ & 1.21             & / 76 & 9.9       & $_{-0.6}^{+0.7}$ & 10.1 & $_{-0.6}^{+0.7}$ \\
P0.65--0.80                                    & 732   & $0.8$    & $_{-0.1}^{+0.1}$ & 0.2 & $_{-0.0}^{+0.4}$ & 1.02             & / 29 & 5.4       & $_{-0.7}^{+1.0}$ & 5.6  & $_{-0.7}^{+1.1}$ \\
P0.80--0.05                                    & 2570  & $1.0$    & $_{-0.1}^{+0.1}$ & 0.7 & $_{-0.2}^{+0.2}$ & 1.03             & / 69 & 11.1      & $_{-0.9}^{+1.0}$ & 12.1 & $_{-1.0}^{+1.1}$ \\
\hline
\end{tabular*}

%% file: tab10.tex
\begin{tabular*}{\textwidth}{c c r@{}l r@{}l  r@{\ }l r@{}l r@{}l}
\hline\hline
   \multicolumn{1}{c}{(1)}                     & (2)   & \multicolumn{2}{c}{(3)     } & \multicolumn{2}{c}{(4)} & \multicolumn{2}{c}{(5)             } & \multicolumn{2}{c}{(6)      } & \multicolumn{2}{c}{(7) } \\
Data                                           & Net   & \multicolumn{2}{c}{$\Gamma$} & \multicolumn{2}{c}{\nH} & \multicolumn{2}{c}{$\chi^2_r$ / DoF} & \multicolumn{4}{c}{$L$\Ss{X}} \\
\cmidrule(r){9-12}Segment                      & Count & \multicolumn{2}{c}{        } & \multicolumn{2}{c}{   } & \multicolumn{2}{c}{                } & \multicolumn{2}{c}{Obs.     } & \multicolumn{2}{c}{Int.} \\
\hline
\multicolumn{1}{l}{\it All}                    & 2292  & $1.2$    & $_{-0.1}^{+0.1}$ & 1.0 & $_{-0.1}^{+0.1}$ & 1.06             & / 47 & 1.9       & $_{-0.1}^{+0.2}$ & 2.2  & $_{-0.2}^{+0.2}$ \\
\hline\multicolumn{3}{l}{\it By folded phases} &                  &     &                  &                  &      &           &                  &      &                  \\
P0.05--0.55                                    & 1459  & $1.3$    & $_{-0.1}^{+0.1}$ & 1.4 & $_{-0.2}^{+0.2}$ & 1.12             & / 42 & 2.2       & $_{-0.2}^{+0.2}$ & 2.7  & $_{-0.3}^{+0.3}$ \\
P0.60--1.00                                    & 668   & $0.8$    & $_{-0.1}^{+0.1}$ & 0.5 & $_{-0.2}^{+0.3}$ & 0.87             & / 27 & 1.5       & $_{-0.2}^{+0.2}$ & 1.6  & $_{-0.2}^{+0.3}$ \\
\hline
\end{tabular*}

%% file: tab11.tex
\begin{tabular*}{\textwidth}{c c r@{}l r@{}l  r@{\ }l r@{}l r@{}l}
\hline\hline
   \multicolumn{1}{c}{(1)}                     & (2)   & \multicolumn{2}{c}{(3)     } & \multicolumn{2}{c}{(4)} & \multicolumn{2}{c}{(5)             } & \multicolumn{2}{c}{(6)      } & \multicolumn{2}{c}{(7) } \\
Data                                           & Net   & \multicolumn{2}{c}{$\Gamma$} & \multicolumn{2}{c}{\nH} & \multicolumn{2}{c}{$\chi^2_r$ / DoF} & \multicolumn{4}{c}{$L$\Ss{X}} \\
\cmidrule(r){9-12}Segment                      & Count & \multicolumn{2}{c}{        } & \multicolumn{2}{c}{   } & \multicolumn{2}{c}{                } & \multicolumn{2}{c}{Obs.     } & \multicolumn{2}{c}{Int.} \\
\hline
\multicolumn{1}{l}{\it All}                    & 4211  & $0.8$    & $_{-0.1}^{+0.1}$ & 1.1 & $_{-0.2}^{+0.2}$ & 0.60             & / 71 & 4.7       & $_{-0.3}^{+0.3}$ & 5.2  & $_{-0.3}^{+0.4}$ \\
\hline\multicolumn{3}{l}{\it By folded phases} &                  &     &                  &                  &      &           &                  &      &                  \\
P0.45--0.70                                    & 1360  & $0.8$    & $_{-0.1}^{+0.1}$ & 1.8 & $_{-0.5}^{+0.5}$ & 1.12             & / 27 & 6.2       & $_{-0.9}^{+1.1}$ & 7.0  & $_{-1.0}^{+1.2}$ \\
P0.70--0.45                                    & 2851  & $0.8$    & $_{-0.1}^{+0.1}$ & 0.8 & $_{-0.2}^{+0.2}$ & 0.98             & / 75 & 4.2       & $_{-0.3}^{+0.4}$ & 4.5  & $_{-0.4}^{+0.4}$ \\
\hline
\end{tabular*}